\documentclass[11pt]{article}
\pdfoutput=1
\usepackage{jheppub}
\usepackage{subfig}
\usepackage{comment}
\usepackage{array,setspace,mathrsfs,amsfonts,amsmath,yfonts,dsfont,bbm,colonequals}

\newcommand\hare[1]{\textcolor{red}{(hare: #1)}}
\newcommand\arash[1]{\textcolor{blue}{(arash: #1)}}

\title{Topologically charged BPS microstates in AdS$_3$/CFT$_2$}

\date{}

\begin{document}

\author{ Arash Arabi Ardehali$\,^*$}
\author{and Hare Krishna$\,^*$${}^\dagger$}

\affiliation{* C.N. Yang Institute for Theoretical Physics,\\ ${}\ $ SUNY, Stony Brook, NY 11794, USA\\
${}\,\dagger$ Weinberg \, Institute,\\
${}\ $ The University of Texas at Austin, TX 78712, USA}

\emailAdd{a.a.ardehali@gmail.com}\emailAdd{hkrishna.phy@gmail.com}

\abstract{In the standard $\mathcal{N}=(4,4)$ AdS$_3$/CFT$_2$ with $\mathrm{sym}^N(T^4)$, as well as the $\mathcal{N}=(2,2)$ Datta-Eberhardt-Gaberdiel variant with $\mathrm{sym}^N(T^4/\mathbb{Z}_2)$, supersymmetric index techniques have not been applied so far to the CFT states with target-space momentum or winding. We clarify that the difficulty lies in a central extension of the SUSY algebra in the momentum and winding sectors, analogous to the central extension on the Coulomb branch of 4d $\mathcal{N}=2$ gauge theories. We define modified helicity-trace indices tailored to the momentum and winding sectors, and use them for microstate counting of the corresponding bulk black holes. In the $\mathcal{N}=(4,4)$ case we reproduce the microstate matching of Larsen and Martinec. In the $\mathcal{N}=(2,2)$ case we resolve a previous mismatch with the Bekenstein-Hawking formula encountered in the topologically trivial sector by going to certain winding sectors.}

\maketitle

%%%%%%
\section{Introduction}

Semi-classical considerations imply that black holes carry entropy \cite{Bekenstein:1973ur,Hawking:1975vcx}
\begin{eqnarray}
    S= \frac{\text{Area}(\mathrm{horizon})}{4 G_N}.
\end{eqnarray}
Statistical derivation of this formula through Boltzmann's relation $S=\log\Omega$ requires a model for the underlying microstates $\Omega$. In a landmark paper, Strominger and Vafa \cite{Strominger:1996sh} provided such a microscopic model for certain five-dimensional BPS black holes in string theory using D-branes \cite{Polchinski:1995mt}. The near-horizon geometry of these black holes involves the three-dimensional BTZ black hole \cite{Banados:1992wn}, therefore in modern parlance their work can be considered an instance of black hole microstate counting in AdS$_3$/CFT$_2$ \cite{Maldacena:1997re}.

The Strominger-Vafa setup consists of a D1-D5 system wrapping an $S^1$, as well as an $M_4$ in the case of the D5 branes, in type IIB string theory. This system has an AdS$_3$ in its near-horizon geometry ($\sim\mathrm{AdS}_3\times S^3\times M_4$), and admits a CFT$_2$ dual in the form of a symmetric orbifold with seed target-space $M_4$. Addition of D-brane momentum along the $S^1$ direction excites the brane system to a black hole, and the  near-horizon AdS$_3$ to a BTZ. The BTZ microstates can then be counted in the CFT$_2$, and in particular those in the BPS sector can be counted via a 2d superconformal index \cite{Maldacena:1999bp,Aharony:1999ti,David:2002wn}.

In the present work we consider the standard $\mathcal{N}=(4,4)$ case where $M_4=T^4$ \cite{Maldacena:1999bp}, and its $\mathcal{N}=(2,2)$ variants discussed in \cite{Datta:2017ert,Eberhardt:2017uup,ArabiArdehali:2021iwe,ArabiArdehali:2023kar} (see also \cite{ArabiArdehali:2018mil,Gaberdiel:2019wjw}). In these cases, the usual 2d superconformal index, namely the elliptic genus, vanishes due to target-space fermionic zero-modes. One must therefore consider \emph{modified} (or \emph{helicity-trace}) \emph{indices}, involving additional fermion-number insertions inside the trace to soak up the zero-modes. This was done in the $M_4=T^4$ case by Maldacena-Moore-Strominger (MMS) \cite{Maldacena:1999bp}, and in the $\mathcal{N}=(2,2)$ variants involving $\mathbb{Z}_k$ orbifolds of $T^4$ in \cite{ArabiArdehali:2023kar}. While MMS found the modified index to reproduce the entropy of the bulk rotating BTZ black holes, the modified indices in \cite{ArabiArdehali:2023kar} reproduced only a fraction $1-\frac{1}{k}$ of the bulk entropy. This mismatch motivated the present work.

The works \cite{Maldacena:1999bp,ArabiArdehali:2023kar} considered only the states without additional charges. In an attempt to shed light on the mismatch found in \cite{ArabiArdehali:2023kar}, here we explore the states carrying charges associated with the momenta and windings on $T^4$ (or its orbifolds). We refer to these as \emph{topologically charged states}. In the $T^4$ case, the microstate counting of such topologically charged black holes was first discussed by Larsen and Martinec \cite{Larsen:1999uk}, but without using supersymmetric indices. Our approach here involves a natural extension of the MMS work \cite{Maldacena:1999bp} to the topological sectors, and thus complements that of Larsen and Martinec \cite{Larsen:1999uk}.

In most of this paper we develop the (modified) SUSY index approach to the topologically charged states in the standard $T^4$ case, reproducing the results of \cite{Larsen:1999uk} via a simple extension of \cite{Maldacena:1999bp}. Section~\ref{sec:2comma2} then applies the machinery to the simplest of the $\mathcal{N}=(2,2)$ cases where a mismatch was found in \cite{ArabiArdehali:2023kar}, and resolves the mismatch in certain charged sectors.

A conservative lesson to take from this analysis would be that the mismatch encountered in \cite{ArabiArdehali:2023kar} arose likely due to large cancellations in the topologically trivial sectors; large accidental cancellations that are absent in other sectors with nonzero topological charge studied here. That large accidental cancellations in SUSY indices may underlie black hole microstate counting mismatches was also suggested in \cite{Vafa:1997gr}.

In the rest of this section we briefly review the bulk/boundary mismatch of \cite{ArabiArdehali:2023kar}, in the simplest and best understood case (with $k=2$), which we refer to as the $\mathrm{HS}_2$ duality. We arrive at the mismatch through a ``modularity argument'', which highlights the significance of the topologically charged states. This motivates our analysis of these states in the BPS sector, which we present in the following sections, first for the standard $\mathcal{N}=4$ symmetric orbifold with seed $T^4$, and then in Section~\ref{sec:2comma2} for the $\mathcal{N}=2$ symmetric orbifold with the hyperelliptic seed $\mathrm{HS}_2=T^4/\mathbb{Z}_2$.

\subsection*{The $\mathrm{HS}_2$ index in the uncharged sector, its modularity and growth}

The setup is the $\mathcal{N}=(2,2)$ AdS$_3$/CFT$_2$ duality between type IIB string theory on AdS$_3\times (S^3\times T^4)/\mathbb{Z}_2$, and the symmetric orbifold CFT$_2$ with target $\mathrm{sym}^N \big(T^4/\mathbb{Z}_2\big)$. The $\mathbb{Z}_2$ acts on $T^4$ by reflecting two of the circles and shifting the other two, while on the bulk $S^3$ ($\subset \mathbb{C}^2$) it acts as $\mathrm{diag}(1,-1)$ \cite{Datta:2017ert,Eberhardt:2017uup}.

We are interested in the SUSY index of the symmetric orbifold CFT. However, let us begin with the index of the seed $\mathrm{HS}_2=T^4/\mathbb{Z}_2$. By the supersymmetric index we here mean the \emph{first helicity-trace index} defined as (see \cite{ArabiArdehali:2023kar})
\begin{equation}
    \mathcal{E}^\text{neut}_1(z,\tau):=\partial_{\bar y} Z^\text{neut}(q,y,\bar q,\bar y)\big|_{\bar y=1}\,,\label{eq:E1_neut}
\end{equation}
with $Z^\text{neut}$ obtained by suppressing the contributions of the topologically charged states to
\begin{equation}
    Z(q,y,\bar q,\bar y)=\mathrm{Tr}_{RR}(-1)^F q^{L_0-c/24} \bar q^{\bar L_0-c/24} y^{2J_0} \bar y^{2\bar J_0}.
\end{equation}
Here $L_0,\bar L_0,J_0,\bar J_0$ are the bosonic quantum numbers of the $\mathcal{N}=(2,2)$ superconformal algebra, and $F:=2(J_0-\bar J_0)$ is our fermion number. The helicity insertion (or $\partial_{\bar y}\big|_{\bar y=1}$) in \eqref{eq:E1_neut} soaks up the fermionic zero-mode of the $\mathrm{HS}_2$ target. The index $\mathcal{E}_1$ was studied in detail in \cite{ArabiArdehali:2023kar} (paralleling MMS \cite{Maldacena:1999bp} who considered the $\mathcal{N}\!=\!4$ \ $T^4$ CFT and studied its \emph{2nd} helicity-trace index to soak up the \emph{two} fermionic zero-modes of $T^4$). 

Denote the index of the seed $\mathrm{HS}_2$ by $\mathcal{H}_1(z,\tau)$. It turns out that $\mathcal{H}_1(z,\tau/2)$ is a weak Jacobi form of weight $-1$ and index $2$ \cite{ArabiArdehali:2023kar}:
\begin{equation}
    \mathcal{H}_1(z,\tau/2)=-i\,\phi_{-1,2}(\tau,z)=-2i\,\frac{\theta_1(z,\tau)}{\eta^3}\frac{\theta_2(z,\tau)}{\theta_2(\tau)}=y^{-1}-y+\mathcal{O}\big(q^{2}\big).\label{eq:HS2wjf}
\end{equation}
That is to say, it enjoys the modular transformation property
\begin{equation}
    e^{-2\pi i z^2/\tau}\,\tau\ \mathcal{H}_1(z/\tau,-1/2\tau)=\mathcal{H}_1(z,\tau/2).\label{eq:HS2modularity}
\end{equation}
Combined with the low-temperature result $\mathcal{H}_1(z,\tau)=y^{-1}-y+\mathcal{O}(q),$ we deduce\footnote{For $z\in(1/2,1)$ a different expression applies \cite{ArabiArdehali:2023kar}, whose derivation via modularity require knowledge of the $\mathcal{O}(q)$ term in the low-temperature expansion of $\mathcal{H}_1.$}
\begin{equation}
\mathcal{H}_1(z,\tau)\xrightarrow{\tau\to0}\tau\, \exp[\frac{i\pi}{\tau}(-2z^2+z)],\quad z\in(0,1/2).\label{eq:H1asy}
\end{equation}

Now, according to DMVV \cite{Dijkgraaf:1996xw}, the---fermion-number weighted---degeneracy $\tilde d(n,N,j)$ of the states with $\Delta=n$, $\ell=j$ encoded in the index $\mathcal{E}^\text{neut}_1\big[\mathrm{sym}^N \mathrm{HS}_2\big]$ of the symmetric orbifold, is (modulo number-theoretic subtleties discussed in Appendix~D of \cite{ArabiArdehali:2023kar}) equal to $\hat{c}_1(N n,j)$. Here $\hat{c}_1(n,j)$ is the Fourier coefficient of $\mathcal{H}_1$. Thus, the degeneracies encoded in the index $\mathcal{E}_1$ can be extracted via a saddle-point analysis as follows:
\begin{equation}
\tilde{d}(n,N,j)\overset{\text{DMVV}\,\&\,\eqref{eq:H1asy}}{\approx} \int_0^1 \mathrm{d}\tau \int_0^1
\mathrm{d}z\ e^{-2\pi i  Nn\tau-2\pi i j z-2\pi i
\frac{z^2}{\tau}+2\pi i \frac{z}{2\tau}} \ \tau.\label{eq:last}
\end{equation}
Extremizing the integrand we find a saddle point at
\begin{equation}
z_0=\frac{1}{4}-\frac{j}{2}\tau_0,\ \ \quad
\tau_0=\frac{i}{4\sqrt{Nn-\frac{j^2}{4}}}.
\end{equation}
The maximized exponent then turns out to have real part
\begin{equation}
S_\text{index}=2\pi\sqrt{Nn-\frac{j^2}{4}}\cdot\frac{1}{2}.\label{eq:entropyB012}
\end{equation}
The saddle-point entropy also has an imaginary piece $-i\pi j/2$, signalling unobstructed phase oscillations in $\tilde d$.

The above entropy is half of the one implied by Cardy's formula, and half of the Bekenstein-Hawking entropy of the bulk BPS black holes \cite{ArabiArdehali:2023kar}. Therefore, it appears that due to large bose-fermi cancellations, the index encodes far (exponentially, to be specific) fewer protected states than expected.

\subsection*{Significance of the topologically charged states}

The derivation of the mismatch as just described, makes it clear that the peculiar modular property \eqref{eq:HS2modularity} is responsible for the conflict: usual elliptic genera or modified indices that correctly reproduce the Bekenstein-Hawking entropy are weak Jacobi forms of index 1 (for the four-real-dimensional Kahler seed, or of index $N=c/6$ for its sym$^N$), whereas here the index (with $\tau$ rescaled by a factor of two) is a weak Jacobi form (WJF) of index 2.

How does the peculiar modular property \eqref{eq:HS2modularity} arise? We argue that it comes from suppressing the contribution of the topologically charged sectors to $Z$.

The full partition function $Z$ has on general grounds the modular property (see \emph{e.g.}~\cite{Kraus:2006nb})
\begin{equation}
    Z\big(\!-\frac{1}{\tau},\frac{z}{\tau},\frac{\bar z}{\bar\tau}\big)=e^{2\pi i\frac{z^2}{\tau}\frac{c}{6}}e^{-2\pi i\frac{\bar z^2}{\bar \tau}\frac{c}{6}}\,Z(\tau,z,\bar z)\,.\label{eq:Zmodular}
\end{equation}
For example, for the $\mathcal{N}=4$ sigma model with target $T^4$ and $c=6$, we have
\begin{equation}
    Z[T^4]=\Theta^{T^4}\left|\frac{\theta_1(z,\tau)}{\eta^3}\right|^4.\label{eq:ZofT4}
\end{equation}
The transformation property \eqref{eq:Zmodular} can be easily confirmed by separating the modular invariant combination $\Theta^{T^4}/|\eta(\tau)|^8$, and the product of weight 0 index 1 WJFs $|\theta_1(z,\tau)/\eta(\tau)|^4$. 

To get a holomorphic supersymmetric index, MMS \cite{Maldacena:1999bp} restricted to the topologically trivial sector and considered the 2nd helicity-trace index:
\begin{equation}
    \mathcal{E}^{\text{neut}}_2\overset{}{=}\frac{\mathrm{d}^2}{\mathrm{d}\bar y^2}Z^{\text{neut}}|^{}_{\bar y=1}\longrightarrow \phi_{-2,1}\,,\label{eq:E2andZasMMS}
\end{equation}
where $\phi_{-2,1}$ is the weight $-2$ index $1$ WJF associated to the $\mathcal{N}\!=\! 4$ \ $T^4$ CFT frequently cited in the literature. Note that suppression of the topologically charged sectors effectively sets 
\begin{equation}
    \Theta^{T^4}\to1,\label{eq:replacement}
\end{equation}
which in turn modifies the modular property from that of $Z$ in \eqref{eq:Zmodular} to that of $\phi_{-2,1}$: since as mentioned below \eqref{eq:ZofT4}, the combination $\Theta^{T^4}/(\eta(\tau)^4\eta(\bar\tau)^4)$ is modular invariant, and since $\eta$ has weight $1/2$ and index $0,$ the replacement \eqref{eq:replacement} reduces the left and right weights by 2, leaving the index of the WJF intact. It thus yields $\phi_{-2,1}$.\\

In the $\mathcal{N}=2$ cases with target $\mathrm{HS}=T^4/\mathbb{Z}_k$, the restriction to the uncharged (or topologically trivial) sector has a more destructive effect on $Z$, which is not so easy to track as in the $\mathcal{N}=4$ case just described. However, since in these cases we have to suppress $\Theta^{T^2}$ \cite{ArabiArdehali:2023kar}, we expect (possibly vector-valued) weak Jacobi forms of weight $-1,$ and indeed this is what we have for $\mathrm{HS}_2$. More importantly, the fact that $\mathcal{H}_1[\mathrm{HS}_2]$ transforms in the peculiar form \eqref{eq:HS2modularity} (and not as an index-1 WJF as we may have wishfully anticipated) must also be due to the suppression of the topologically charged sectors.

%%%%%%%%%%
\subsection*{Inclusion of the topological sectors}

A quick and dirty way of incorporating the topological sectors is to simply relax the restriction to the neutral sector in \eqref{eq:E1_neut} and \eqref{eq:E2andZasMMS}. For instance, in the $T^4$ case we can consider
\begin{equation}
    \mathcal{E}_2\overset{!}{=}\frac{\mathrm{d}^2}{\mathrm{d}\bar y^2}Z|^{}_{\bar y=1}=(\frac{1}{2\pi i}\frac{\mathrm{d}}{\mathrm{d}\bar z})^2Z|^{}_{\bar z=0}.\label{eq:E2!Z}
\end{equation}
Using the modular property of $Z$ as in \eqref{eq:Zmodular}, we then get
\begin{equation}
    \begin{split}
    \mathcal{E}_2(-1/\tau,z/\tau)&\overset{!}{=}(\frac{1}{2\pi i}\frac{\mathrm{d}}{\mathrm{d}\bar z/\bar\tau})^2 Z|^{}_{\bar z=0}(-1/\tau,z/\tau,\bar z/\bar\tau)\\
    &=\bar\tau^2(\frac{1}{2\pi i}\frac{\mathrm{d}}{\mathrm{d}\bar z})^2\big(e^{2\pi i\frac{c}{6}(z^2/\tau-\bar z^2/\bar\tau)}Z(\tau,z,\bar z)\big)\big|_{\bar z=0}=\bar\tau^2\, e^{2\pi i\frac{c}{6}z^2/\tau}\,\mathcal{E}_2(\tau,z)\,.
    \end{split}\label{eq:initialMod}
\end{equation}
This would yield the right asymptotic to match with the Cardy (or Bekenstein-Hawking) entropy.
But as the factor of $\bar\tau^2$ above indicates, the $\mathcal{E}_2$ defined via \eqref{eq:E2!Z} is not holomorphic. Therefore it is unclear whether the states counted as such are protected.

The more methodic approach in the body of this paper illustrates how a minor variation puts \eqref{eq:E2!Z} on the right track. We will leverage the modified, centrally extended, SUSY algebra in the topologically charged sectors, which is analogous to the centrally extended SUSY algebra arising in Coulomb branch physics of 4d $\mathcal{N}=2$ gauge theories \cite{Seiberg:1994rs}. In the $T^4$ case, the corrected variant leads to a somewhat trivial re-organization of the MMS computations in \cite{Maldacena:1999bp}, and reproduces the results of \cite{Larsen:1999uk}. In the $\mathrm{HS}_2$ case, on the other hand, it amounts to more than a re-organization: a $\mathbb{Z}_2$ twisted sector absent in \eqref{eq:E1_neut} ends up playing the key role, and restores the match with the Bekenstein-Hawking entropy.

%%%%%%%%%%
\subsection*{Outline of the paper}

The rest of this article is structured as follows. In Section \ref{sec:T4_alg}, we discuss the modified (centrally extended) supersymmetric algebra in the topologically charged sectors of the $\mathcal{N}=(4,4)$ $T^4$ sigma model. The SUSY algebra provides a shortening condition which is discussed in Section~\ref{sec:shortening_T4}, where we also compare with the shortening condition arsing in the large $\mathcal{N}=4$ algebra. We moreover explain how to go from the shortening condition and the BPS bound in $T^4$, to those in $\mathrm{sym}^N (T^4)$. In Section~\ref{sec:pn_fns_T4},  we discuss the supersymmetric index that encodes the topologically charged BPS states in the $\mathcal{N}=(4,4)$ sigma models with target $T^4$ or (via the DMVV formula \cite{Dijkgraaf:1996xw}) $\mathrm{sym}^N (T^4)$. In Section~\ref{sec:saddle_point_T4}, we compute the microscopic entropy via a saddle-point analysis of the index of $\mathrm{sym}^N (T^4)$, reproducing the Bekenstein-Hawking entropy of the charged black holes in the bulk, corroborating the Larsen-Martinec derivation \cite{Larsen:1999uk}. Section~\ref{sec:2comma2} extends the analysis to the $\mathcal{N}=(2,2)$ case of $\mathrm{HS}_2$ and $\mathrm{sym}^N(\mathrm{HS}_2).$ We finish the article with a discussion of future directions in Section~\ref{sec:discussion}. Some technical details regarding the DMVV formula with the inclusion of the topological charges, and the derivation of the topological contributions to the $\mathrm{HS}_2$ partition function, are relegated to the appendices.

\section{Modified SUSY algebra in the seed topological sectors}
\label{sec:T4_alg}

In this section we study the $U(1)^8$ extended $\mathcal{N}=(4,4)$ superconformal algebra governing $\mathcal{N}=(4,4)$ supersymmetric sigma models such as those with $T^4$ or $\mathrm{sym}^N(T^4)$ target space. The $U(1)^8$ extension is due to the momentum and winding sectors of each of the four circles (or their overall-acting counterparts in the case of $\mathrm{sym}^N(T^4)$). We focus on the case of $T^4$, and explain in the following sections how the corresponding results for $\mathrm{sym}^N(T^4)$ can be obtained.

In the conventions of \cite{Hampton:2018ygz}, the four real right-moving fermions $\psi_i, \quad i=1,2,3,4$ of the $T^4$ sigma model can be written as doublets $\psi^{\alpha A}$:
\begin{eqnarray}
    (\psi^{++},\psi^{-+})&&= \frac{1}{\sqrt{2}} (\psi_1+i \psi_2, \psi_3+i \psi_4),\\
     (\psi^{+-},\psi^{--})&&= \frac{1}{\sqrt{2}} (\psi_3-i \psi_4, -\psi_1+i \psi_2).
\end{eqnarray}
The index $\alpha=(+,-)$ corresponds to the $R$-symmetry group $SU(2)_R$, while the index $A=(+,-)$ corresponds to the subgroup $SU(2)$ of the rotations of the tangent space of $T^4$.  Similarly, the 4 real right-moving bosons $X_1,X_2,X_3,X_4$ can be grouped into
\begin{eqnarray}
    X_{A\dot{A}}= \frac{1}{\sqrt{2}} X_i \sigma_i =  
  \frac{1}{\sqrt{2}} \left(
\begin{array}{cc}
X_3+i X_4 & X_1-i X_2  \\
 X_1+i X_2 & -X_3+i X_4 \\
\end{array}
\right).   
 \end{eqnarray}
Here $\sigma_i= (\sigma_a,i I)$.

We consider the right-handed algebra, which is a $U(1)^4$ extended $\mathcal{N}=4$ superconformal algebra. The following rescaling of the right-handed supercharges allows us to go from the (anti-)commutators obtained in \cite{Hampton:2018ygz} to those of MMS \cite{Maldacena:1999bp}:
\begin{eqnarray}
    \bar{G}_0^{+-} \rightarrow i \sqrt{2}  \bar{G}_0^{+-} ,&\quad    \bar{G}_0^{-+} \rightarrow i \sqrt{2}  \bar{G}_0^{-+} ,\nonumber\\
        \bar{G}_0^{++} \rightarrow  \sqrt{2}  \bar{G}_0^{++} ,&\quad    \bar{G}_0^{--} \rightarrow  \sqrt{2}  \bar{G}_0^{--},\nonumber\\
        \bar{Q}_0^{+-} \rightarrow i   \bar{Q}_0^{+-} ,&\quad    \bar{Q}_0^{-+} \rightarrow i   \bar{Q}_0^{-+} .\nonumber\label{eq:T4_alg_modified}
\end{eqnarray}
With this rescaling, the Ramond sector zero-mode algebra of  the right-handed sector can be written as 
  \begin{align}
 \label{eq:zero_mode_susy_T4_main}
    \lbrace \bar{G}_0^{++},\bar{G}_0^{--}\rbrace&= 2 \bar{L}_0, &  \lbrace\bar{G}_0^{+-},\bar{G}_0^{-+}\rbrace&= 2 \bar{L}_0,\nonumber\\
    \lbrace \bar{Q}_0^{++},\bar{Q}_0^{--}\rbrace&= 1,& \lbrace \bar{Q}_0^{+-},\bar{Q}_0^{-+}\rbrace&= 1,\nonumber\\
   \lbrace \bar{G}_0^{++},\bar{Q}_0^{--}\rbrace&= -\frac{i}{2}(\bar u_1+i \bar u_2),& \lbrace \bar{G}_0^{++},\bar{Q}_0^{-+}\rbrace&= \frac{1}{2}(\bar u_3+i \bar u_4),\nonumber\\
     \lbrace \bar{G}_0^{+-},\bar{Q}_0^{--}\rbrace&= -\frac{i}{2}(\bar u_1+i \bar u_2),& \lbrace \bar{G}_0^{+-},\bar{Q}_0^{-+}\rbrace&= \frac{1}{2}(\bar u_3+i \bar u_4),\nonumber\\
     [\bar{J}_0^3,\bar{G}_0^{\pm-}]&=\pm \frac{1}{2} \bar{G}_0^{\pm-},& [\bar{J}_0^3,\bar{G}_0^{\pm+}]&=\pm \frac{1}{2} \bar{G}_0^{\pm+},\nonumber\\
      [\bar{J}_0^3,\bar{Q}_0^{\pm-}]&=\pm \frac{1}{2} \bar{Q}_0^{\pm-},& \ [\bar{J}_0^3,\bar{Q}_0^{\pm+}]&=\pm \frac{1}{2} \bar{Q}_0^{\pm+}.
 \end{align}
Here $\bar G_0^{\pm \pm}$ are the supercharges, with the first $\pm$ superscript indicating the charge under $\bar J_0^3$ of $\mathrm{SU}(2)_R$.  The operators $\bar Q_0^{\pm\pm}$ are the fermionic partners of the $U(1)^4$ current algebra. The chiral charges
\begin{equation}
    \bar u_i= \frac{m_i}{R_i}- w_i R_i\,,
\end{equation}
are real. Here $m_i$ and $w_i$ correspond to momentum and winding charges of a compact boson with radius $R_i$. The algebra in the zero momentum and winding sectors is obtained simply by setting $\bar u_i=0$ (see  Eq.~(3.9) of \cite{Maldacena:1999bp}). The most non-trivial part of the algebra is the anti-commutators of $\bar G$ and $\bar{Q}$ in the non-zero momentum and winding sectors, which is derived in a similar but simpler setting in Section~\ref{subsec:2comma2_alg}. The algebra with nonzero $\bar u_i$ can be thought of as a \emph{central extension} of the SUSY algebra with all $\bar u_i=0.$

The algebra can be written as a fermionic mode algebra if we identify as fermion creation operators (denote them $b^{\dagger}_i$) the generators $\bar G_0^{++},\,\bar G_0^{+-},\,\bar Q_0^{++},\,\bar Q_0^{+-}$, and as annihilation operators (denote them $b_i$) the generators $\bar G_0^{--},\,\bar G_0^{-+},\,\bar Q_0^{--},\,\bar Q_0^{-+}$. Then the first four lines of the SUSY algebra \eqref{eq:zero_mode_susy_T4_main} reduce to
 \begin{eqnarray}
\{b_i,b_j\}=0, \quad \{b_i^{\dagger}, b_j^{\dagger}\}=0,\quad \{b_i,b_j^{\dagger}\}=M_{ij}\,.
\end{eqnarray}
The matrix $M_{ij}$ can be written as
\begin{eqnarray}
   M_{ij}=\left(
\begin{array}{cccc}
 2 \bar{L}_0 & 0 & -\frac{i}{2} (\bar u_1+i \bar u_2) & \frac{1}{2} (\bar u_3+i \bar u_4) \\
 0 & 2 \bar{L}_0 & -\frac{i}{2} (\bar u_1+i \bar u_2) & \frac{1}{2} (\bar u_3+i \bar u_4) \\
 \frac{i}{2} (\bar u_1-i \bar u_2) & \frac{i}{2} (\bar u_1-i \bar u_2) & 1 & 0 \\
 \frac{1}{2} (\bar u_3-i \bar u_4) & \frac{1}{2} (\bar u_3-i \bar u_4) & 0 & 1 \\
\end{array}
\right)   .
 \end{eqnarray}

The convention for winding and momentum that we used is the usual convention often used in string theory (see Polchinski 8.2.4 \cite{Polchinski:1998rq} with $\alpha'=1$)
\begin{eqnarray}
    \partial X(z)= - i\,(\frac{1}{2})^{1/2} \sum_{m=-\infty}^{\infty} \frac{\alpha_m}{z^{m+1}}\,, \quad \alpha_0= \frac{1}{\sqrt{2}}(\frac{n}{R}+w R)\,.
\end{eqnarray}

All other non-trivial anti-commutators (in the right-handed Ramond sector) can be found from \eqref{eq:T4_alg_modified} via the Hermitian conjugation:
\begin{equation}
    (G_0^{\pm \pm})^{\dagger}=G_0^{\mp \mp} , \quad (Q_0^{\pm \pm})^{\dagger}=Q_0^{\mp \mp}.
\end{equation}

%%%%%%%
\section{Shortening condition and BPS states}
\label{sec:shortening_T4}

We begin in this section with the SUSY algebra in the Ramond sector with the inclusion of the topological states (called ``charged states'' throughout the article). With their inclusion, we will see that there is a BPS bound on the right movers given by $\bar{L}_0\ge \frac{1}{4}\sum_{i=1}^4(\bar u_i^2) $, where $\bar u_i:=( \frac{m_i}{R_i}- w_i R_i)$ are the ``anti-holomorphic'' combinations of the momenta $m_i$ and winding $w_i$ quantum numbers, which can be thought of as the chiral charges of the right-handed $U(1)^4$ algebra. We will define an index that counts states satisfying $\bar{L}_0= \frac{1}{4}\sum_{i=1}^4(\bar u_i^2)$, and vanishes on the long representations (of right movers). This index improves on the 2nd helicity-trace index introduced in \cite{Maldacena:1999bp} which counted only states with $\bar{L}_0=0$.\\

We can diagonalize the $M_{ij}$ matrix and find the eigenvalues
\begin{eqnarray}
   1,\ 2 \bar{L}_0,\ \frac{1}{2} \left(-\sqrt{4 \bar{L}_0^2-4 \bar{L}_0+2 \bar u_1^2+2 \bar u_2^2+2 \bar u_3^2+2 \bar u_4^2+1}+2
  \bar{ L}_0+1\right),\nonumber\\
   \frac{1}{2} \left(\sqrt{4 \bar{L}_0^2-4 \bar{L}_0+2 \bar u_1^2+2 \bar u_2^2+2 \bar u_3^2+2 \bar u_4^2+1}+2
  \bar{ L}_0+1\right).
\end{eqnarray}
We see that when 
\begin{equation}
    \bar{L}_0= \frac{1}{4}(\bar u_1^2+\bar u_2^2+\bar u_3^2+\bar u_4^2),\label{eq:U1^4_ext_N=4_BPS}
\end{equation}
one of the eigenvalues becomes zero. We thus get a ``shortening'' condition.
%\footnote{It is interesting to understand the consequence of unitarity for the states which does not saturate this bound. \arash{Can you clarify?}}
%To compare with the previous study of the $T^4$ case in \cite{Maldacena:1999bp}, note that they set $\bar{L}_0=0$ and show that the second helicity-trace index vanishes on the long representations (of right movers). Hence their index captures the BPS states with $\bar{L}_0=0$ on the right-movers side. We will define a similar second helicity-trace index, and with the same logic the long multiplets on right movers give zero contribution, but we will be capturing $\bar{L}_0\neq0$ states as well.\footnote{To argue that the quantum numbers under the left algebra are protected we need a little more: $i$) $L_0-\bar L_0\in\mathbb{Z},$ and $ii$) quantization of $J_0$ charge to justify protectedness.}

The states satisfying \eqref{eq:U1^4_ext_N=4_BPS} in the $\mathcal{N}=(4,4)$ $T^4$ sigma model are (right-handed) BPS. When at least one $\bar u_j$ is nonzero, we will refer to these BPS states as \emph{(topologically) charged BPS states}.

\begin{comment}
\arash{N=4 analog of $\big(\bar G^-_0-\frac{u_2+iu_1}{\sqrt{2}}\bar Q^-_0\big)\big|\bar h,u_1,u_2\rangle.$ ... probably it is:
$$\big(\underbrace{\bar G^{--}_0-\frac{u_2+iu_1}{\sqrt{2}}\bar Q^{--}_0-i\frac{u_3+iu_4}{\sqrt{2}}\bar Q^{-+}_0}_{\text{supercharge underlying the index}}\big)\big|\bar h,u_1,u_2,u_3,u_4\rangle.$$}
\end{comment}

For a $U$-duality perspective on such BPS states see \cite{Dijkgraaf:1996cv,Dijkgraaf:1996hk}.

%%%%%
\subsection{Side note: connection with the large $\mathcal{N}=4$ algebra}

The large $\mathcal{N}=4$ algebra has bosonic subalgebra $\mathfrak{su}(2)_{k_+}\times \mathfrak{su}(2)_{k_-}\times \mathfrak{u}(1)_{\frac{k_++k_-}{2}}$ \cite{Gaberdiel:2013vva}. Its BPS states satisfy
\begin{equation}
    h\ge\frac{1}{k_++k_-}\Big[k_+\ell_-+k_-\ell_++u^2+(\ell_+-\ell_-)^2\Big]\,.\label{eq:large_N=4_BPS}
\end{equation}

In the $k_-\to\infty$ limit, the group manifold $\mathrm{SU(2)}$ decompactifies, and we expect to recover from $\mathfrak{su}(2)_{k_-}\times \mathfrak{u}(1)_{\frac{k_++k_-}{2}}$ part of our $U(1)^4$ extended small $\mathcal{N}=4$ algebra \cite{Sevrin:1988ew}. Keeping $k_+$ fixed and setting $\ell_+=0$ to focus on the $\mathfrak{su}(2)_{k_-}\times \mathfrak{u}(1)_{\frac{k_++k_-}{2}}$ sector, we find from \eqref{eq:large_N=4_BPS}:
\begin{equation}
    h\ge\frac{1}{k_-}\Big[u^2+\ell_x^2+\ell_y^2+\ell_z^2\Big]\,,\label{eq:large_N=4_BPS_contracted0}
\end{equation}
where we have written $\ell_-^2=\ell_x^2+\ell_y^2+\ell_z^2.$ Rescaling $u,\ell_{x,y,z}\to\sqrt{\frac{k_-}{2}}u,\sqrt{\frac{k_-}{2}}\ell_{x,y,z}$, we have
%\arash{instead of this scaling, we can just compare the case $k_-=2,\ k_+=0$ with the Sugawara result below} \hare{in the eq 3.6, you can set $k_-=2,\ell_+=0$ and get the sugawara answer. But the group $su(2)$ is not decompactified yet. }
\begin{equation}
    h\ge\frac{1}{2}\Big[u^2+\ell_x^2+\ell_y^2+\ell_z^2\Big]\,,\label{eq:large_N=4_BPS_contracted}
\end{equation}
which is compatible with \eqref{eq:U1^4_ext_N=4_BPS}, up to a change of normalization.

%\arash{Fixing the factor of 2 compared with \eqref{eq:U1^4_ext_N=4_BPS}...} \hare{factor of 4 instead of 2 all depends on the choice of normalization of generators. To compare these answers, one needs to check their susy algebras.}

%\arash{Comparison of the nulls doesn't seem to make that much sense, because only $\ell_z$ appears in the null vector of the large algebra \cite{Gaberdiel:2013vva}...}

%%%%%
\subsection{From the seed to the symmetric orbifold}

In the seed (supersymmetric) $T^4$ sigma model, every state lacking right-handed oscillators is a BPS state. These lead to a diverse set of BPS states in various twisted sectors of the symmetric orbifold CFT with target $(T^4)^N/S_N\,.$ An efficient approach to understanding the symmetric orbifold BPS states is through partition functions and the DMVV formula, which will be discussed in the next section. Here we only discuss the BPS bound in the topologically charged sectors of the symmetric orbifold.

It is easiest to see the BPS bound in the symmetric orbifold by writing (see \emph{e.g.}~\cite{Aharony:2024fid}):
\begin{equation}
    (T^4)^N/S_N\,=\,\big((T^4)^N/S_N\big)/U(1)_N^8\rtimes  U(1)_N^8\,.
\end{equation}
Here the level of the AKM algebra $U(1)^8$ is denoted as subscript. The above rewriting separates the oscillators from the topologically charged modes of the symmetric orbifold. The right-handed energy carried by the topological modes is given by the Sugawara energy
\begin{equation}
    E^R_\text{Sugawara}= \frac{1}{4N}(\bar u_1^2+\bar u_2^2+\bar u_3^2+\bar u_4^2)\,.\label{eq:sugawara}
\end{equation}
Turning on left-handed oscillators will not change the right-handed energy. So the BPS bound takes the form
\begin{equation}
    \bar L_0\ge \frac{1}{4N}(\bar u_1^2+\bar u_2^2+\bar u_3^2+\bar u_4^2)\,.\label{eq:sugawara_bound}
\end{equation}

To see which twisted sector of the symmetric orbifold realizes the bound, consider the symmetric orbifold Hilbert space  structure \cite{Dijkgraaf:1996xw}:
\begin{equation}
    \mathcal{H}=\bigoplus_{[g]}\mathcal{H}_g^{C_g}=\bigoplus_{[g]}\bigotimes_{\substack{n>0}} S^{N_{n}} \mathcal{H}^{\mathbb Z_n}_{(n)}.
\end{equation}
Here $\mathcal{H}^{\mathbb Z_n}_{(n)}$ stands for the $\mathbb Z_n$ invariant subsector of the Hilbert space $\mathcal{H}_{(n)}$ of a single string on $T^4 \times S^1$ winding $n$ times around the $S^1$. Alternatively, we have a long string of length $2\pi n$ for $\mathcal{H}_{(n)}$, on which $\mathbb Z_n$ acts via a $2\pi$ shift. The $N_n$ are subject to $\sum_n n N_n=N.$

It is straightforward now to see that the bound \eqref{eq:sugawara_bound} arises in what may be called the maximally twisted sector, with $N_N=1.$  The $N$ in the denominator of \eqref{eq:sugawara_bound} arises in this picture due to the length of the long string being $2\pi N.$ 

Note that while the twist operator has dimension $\frac{c_{T^4}}{24}\big(N-\frac{1}{N}\big)=\mathcal{O}(N)$ in the NS sector, our computations are in the Ramond sector and hence our bound is $\mathcal{O}(1/N).$

%%%%
\section{Encoding the BPS states into partition functions}
\label{sec:pn_fns_T4}

\subsection{Defining an index}

The following index vanishes on the long representations of the (right-handed) $U(1)^4$ extended $\mathcal{N}=4$ superconformal algebra, and is therefore protected \cite{Maldacena:1999bp}:
\begin{equation}
    \mathrm{Tr}\,(-1)^{-2\bar{J}_0^{\,3}}(2\bar{J}_{0}^{\,3})^2\,.\label{eq:bare_index}
\end{equation}
However, it needs to be regulated, due to the trace over an infinite-dimensional Hilbert space.

In the topologically trivial sectors, a natural regulator is $q^{L_0}$, associated with the generator $L_0$ in the left-handed algebra. A further refinement using $y^{2J_0^3}$ then gives
\begin{equation}
    \mathcal{E}^{\text{top.$\,$triv.}}_2(q,y)=\mathrm{Tr}_{\text{top.$\,$triv.}}\,(-1)^{F}(2\bar J_0^{\,3})^2q^{L_0}\,y^{2J_0^3},
\end{equation}
with $F=2J_0^3-2\bar J_0^{\,3}$. This is the (helicity-trace) index of Maldacena-Moore-Strominger \cite{Maldacena:1999bp}.

In the topologically non-trivial sectors, labeled by $m_i,\,w_i,$ the refinement with $y^{2J^3_0}$ is still possible. One can even go to a canonical ensemble with respect to the topological charges by introducing associated fugacities $\zeta_i,\,\xi_i$, further refining the index. However, a regulator is still needed.

Introducing $q^{L_0}$ would unfortunately make the index somewhat unconventional. That is because the $L_0$ quantum numbers receive contributions from $m_i/R_i$ and $w_i R_i,$ which would introduce $R_i$-dependence in the index. This is undesirable since $R_i$ are part of the Kahler moduli that conventional indices should not depend on. 

For the seed sigma model, one possible fix would be to introduce $q^{L_0^\text{osc.}}$, with $L_0^\text{osc.}$ the oscillator part of the left-handed energy: 
\begin{equation}
    L_0^\text{osc.}=L_0-\frac{\delta(m_i,w_i)}{4},
\end{equation}
where
\begin{equation}
    \delta(m_i,w_i):=\sum_{i=1}^4 u_i^2,
\end{equation}
with $u_i:=\frac{m_i}{R_i}+w_iR_i\,.$ (We can similarly define $\bar\delta(m_i,w_i):=\sum_{i=1}^4 \bar u_i^2\,$.)

%The index defined as such is \emph{not universal}, in the sense that it relies not only on the underlying super-algebra, but also on the structure of the Hilbert space.

More generally, for the symmetric orbifold $\mathrm{sym}^N(T^4)$, instead of $L_0^\text{osc.}$  we can write\footnote{This can be seen clearly from the DMVV formula discussed below. From \eqref{eq:curlyZ_s} in the sector $s=N/n$ we have $m_i,w_i$ multiplied by $N/n$. So the topological charges are $\frac{N}{n}m_i,\frac{N}{n}w_i.$ On the other hand, the topological excess energy in that sector follows from \eqref{eq:curlyZ_s} to be $\frac{\delta(m_i,w_i)}{4n}\times\frac{N}{n}=\frac{\delta(\frac{N}{n}m_i,\frac{N}{n}w_i)}{4N}$. Therefore indeed removing $\delta/4N$ leaves us with only the oscillator energy.}
\begin{equation}
    L_0-\frac{\delta(m_i,w_i)}{4N},
\end{equation}
compatibly with \eqref{eq:sugawara_bound}.

The combination $L_0-\frac{\delta(m_i,w_i)}{4N}$ commutes with the supercharge underlying the index \eqref{eq:bare_index}, so introducing the fugacity $q^{L_0-\frac{\delta(m_i,w_i)}{4N}}$ is perfectly allowed. Because this involves conserved charges and $N,$ it is slightly better than $q^{L_0^\text{osc.}}$, but still the appearance of $N$ and the specific function $\delta(m_i,w_i)$ make it unfortunately very specific to the $\mathrm{sym}^N(T^4)$ symmetric orbifold sigma model (and the universal character of the index is lost).

Our proposal would then be
\begin{equation}
    \mathcal{E}_2(q,y,\zeta_i,\xi_i)=\mathrm{Tr}\,(-1)^{2J_0^3-2\bar J_0^3}(2\bar J_0^3)^2q^{L_0-\frac{\delta(m_i,w_i)}{4N}}\,y^{2J_0^3}\,\prod_{i=1}^{4}\big(\zeta_i^{m_i}\xi_i^{w_i}\big)\,.
\end{equation}
To keep the notation light, below we will often suppress the products $\prod_{i=1}^4$ such as the one on the right-hand side above.

Alternatively, in the microcanonical ensemble with respect to the topological charges:
\begin{equation}
    \mathcal{E}^{m_i,w_i}_2(q,y)=\mathrm{Tr}_{m_i,w_i}\,(-1)^{2J_0^3-2\bar J_0^3}(2\bar J_0^3)^2q^{L_0-\frac{\delta(m_i,w_i)}{4N}}\,y^{2J_0^3}.\label{eq:mc_index}
\end{equation}

%We emphasize again that this is \emph{not} a universal definition: it is not applicable to all theories with $U(1)^4$-extended $\mathcal{N}=4$ superconformal algebra, but only works for the $\mathcal{N}=4$ $\mathrm{sym}^N(T^4)$ theory of our interest here.

Let us now compute the index $\mathcal{E}_2^{m_i,w_i}$ of the $T^4$ sigma model. We begin with the well-known non-holomorphic partition function
\begin{equation}
    Z(q,y,\bar q,\bar y):=\mathrm{Tr}_{RR}(-1)^F q^{L_0-c/24} \bar q^{\bar L_0-c/24} y^{2J_0} \bar y^{2\bar J_0},\label{eq:Z}
\end{equation}
 given for the $\mathcal{N}=(4,4)$ $T^4$ sigma model by \cite{Maldacena:1999bp}:
\begin{eqnarray}
    Z(q,\bar{q},y,\bar{y})= \sum_{\Gamma^{4,4}} q^{\frac{P_L^2}{4}}\bar{q}^{\frac{P_R^2}{4}} \left|\frac{\theta_1(z,\tau)}{\eta^3}\right|^4\,.
\end{eqnarray}
Here $\Gamma^{4,4}$ is the lattice of the momentum and winding modes. We will pick the simplest lattice, with four perpendicular circles of radii $R_{1,2,3,4}.$
%a particular lattice vector as summarized in the appendix \ref{app:lattices}.
The left and right moving momentum vectors are $P_L^2=\sum_i(\frac{m_i}{R_i}+w_i R_i)^2 =\delta(m_i,w_i)$ and $P_R^2=\sum_i(\frac{m_i}{R_i}-w_i R_i)^2 =\bar\delta(m_i,w_i)$.

To compute the index, we first consider the contribution of a specific topological sector to the partition function: 
\begin{eqnarray}\label{eq:Z_T4}
    Z^{m_i,w_i}(q,\bar q,y,\bar y)=q^{\frac{\sum_i(\frac{m_i}{R_i}+w_i R_i)^2}{4}} \bar {q}^{  \frac{\sum_i(\frac{m_i}{R_i}-w_i R_i)^2}{4}}\left|\frac{\theta_1(z,\tau)}{\eta^3}\right|^4.%\\
   %&&\equiv  Z^{\underline{e_i},\underline{m_i}}(q,\bar{q},y,\bar y) \bar {q}^{  \frac{(\frac{e_i}{R_i}-m_i R_i)^2}{4}}
\end{eqnarray}
%This provides an explicit example of the new partition function that we have talked about. We will restrict to the fixed topological sector $(e_i,m_i)$ and use the partition function $Z^{\underline{e_i},\underline{m_i}}(q,\bar{q},y,\bar y)$ to compute the index.\\

Comparison of the partition function \eqref{eq:Z} and the index \eqref{eq:mc_index} shows that the index can be obtained by setting $\bar q=1$ in $Z$, removing $\delta(m_i,w_i)/4$ from $L_0,$ and finally taking\footnote{The factor of $\frac{1}{2}$ in $\frac{1}{2}\partial^2_{\bar y}\big|_{\bar y=1}$ is inserted to simplify some of the following expressions, but it is completely irrelevant for our purposes.} $\frac{1}{2}\partial^2_{\bar y}\big|_{\bar y=1}$ to implement the insertion of $(2\bar{J}^3_0)^2$. The result is
\begin{equation}
    \mathcal{E}^{m_i,w_i}_2[T^4]=\left(\frac{\theta_1(z,\tau)}{\eta^3}\right)^2.\label{eq:E2_mw_T4}
\end{equation}
In other words, the index is the same irrespective of the topological sector. This is not surprising, since it simply counts the oscillator excitations on top of topological ``vacua'' $|m_i,w_i\rangle$. In particular, it coincides with the MMS index \cite{Maldacena:1999bp} for the excitations on top of $|0,0\rangle$.

\begin{comment}
The BPS bound at the right movers have $\bar{L}_0= \frac{1}{4}\sum_i (\frac{a_i}{R_i}- b_i R_i)^2$. So, for a given $a_i,b_i$  we identify the $a_i=e_i, \, \, b_i=m_i$. So, for any fix $(a_i,b_i)$ saturating the BPS bound, we have corresponding modes in the partition function. Here they are exactly identified to each other.
% \footnote{In HS cases, we will find the $a_i= \frac{e_i}{2}$ and $b_i=m_i$. }
Hence, for a given charged sector, we have
\begin{eqnarray}
Z^{\underline{a_i},\underline{b_i}}(q,\bar{q},y,\bar y)= q^{\frac{1}{4}\sum_i(\frac{a_i}{R_i}+ b_i R_i)^2}  \left|\frac{\theta_1(z,\tau)}{\eta^3}\right|^4\equiv q^{\frac{\delta}{4}}  \left|\frac{\theta_1(z,\tau)}{\eta^3}\right|^4
\end{eqnarray}
Here we have written $\delta=\sum_i(\frac{a_i}{R_i}+ b_i R_i)^2$. 

We can write this partition function as some mode expansion in the oscillator as
\begin{eqnarray}
\label{coeff}
Z^{\underline{a_i},\underline{b_i}}(q,\bar{q},y,\bar y)= \sum_{\Delta,\bar{\Delta},\ell,\bar{\ell}}  c(\Delta,\bar{\Delta},\ell,\bar{\ell}) q^{\Delta} \bar{q}^{\bar{\Delta}} y^{\ell} \bar{y}^{\bar{\ell}} q^{\delta}   
\end{eqnarray}
With this partition function, we will first do the symmetric orbifolds and then calculate the index for symmetric orbifold theories.
\end{comment}

%%%%
\subsection{From the seed to the symmetric orbifold via DMVV}

Writing the seed partition function as
\begin{eqnarray}
\label{coeff}
Z(q,\bar{q},y,\bar y,\zeta_i,\xi_i)= \sum_{\Delta,\bar{\Delta},\ell,\bar{\ell},m_i,w_i}  \!\!c(\Delta,\bar{\Delta},\ell,\bar{\ell},m_i,w_i)\, q^{\Delta+\frac{\delta}{4}} \bar{q}^{\bar{\Delta}+\frac{\bar\delta}{4}} y^{\ell} \bar{y}^{\bar{\ell}}\zeta_i^{m_i}\xi_i^{w_i},
\end{eqnarray}
the DMVV formula gives the symmetric orbifold partition function as (see Appendix~\ref{app:dmvv}):
\begin{equation}
    \mathcal{Z}(p,q,\bar{q},y,\bar{y},\zeta_i,\xi_i)\equiv e^{\mathcal{F}}=  \prod_{n=1}^{\infty} \prod_{\Delta,\bar{\Delta},\ell,\bar{\ell},m_i,w_i}^{'} \frac{1}{\big(1-p^n q^{\frac{\Delta}{n}+\frac{\delta}{4n}}\bar{q}^{\frac{\bar{\Delta}}{n}+\frac{\bar\delta}{4n}}y^{\ell} \bar{y}^{\bar{\ell}}\zeta_i^{m_i}\xi_i^{w_i}\big)^{c(\Delta,\bar{\Delta},\ell,\bar{\ell},m_i,w_i)}}.\label{eq:DMVVmw}
\end{equation}
The prime in the product refers to the restriction $(\Delta- \bar{\Delta}+\sum_i m_i w_i) \in n\, \mathbb{Z}\,.$ In the BPS sector where $\bar\Delta=0,$ this translates to
\begin{equation}
    \Delta+\sum_i m_i w_i \in n\, \mathbb{Z}\,.\label{eq:BPS'}
\end{equation}

%%%%
\subsection*{Explicit calculation for $\mathrm{sym}^N T^4$}

For the $T^4$ sigma model, we also have $Z(X)|_{\bar{y}=1}=Z(X)'|_{\bar{y}=1}=0$. This is due to the $T^4$ fermion zero modes.  Hence we have the following relations for the coefficients
\begin{eqnarray}
  &&  \sum_{\bar{\ell}} c(\Delta,\bar{\Delta},\ell,\bar{\ell},m_i,w_i)=0,\nonumber\nonumber\\
    && \sum_{\bar{\ell}} \bar{\ell} c(\Delta,\bar{\Delta},\ell,\bar{\ell},m_i,w_i)=0,\nonumber\\
    &&  \sum_{\bar{\ell}} \bar{\ell}^2 c(\Delta,\bar{\Delta},\ell,\bar{\ell},m_i,w_i)=0, \quad \text{for }\,\,\, \bar{\Delta}>0 \,.\label{eq:zero_sums}
\end{eqnarray}
These relations can be explicitly checked by expanding the partition function \eqref{eq:Z_T4} of the $T^4$ sigma model and taking the appropriate derivative. With this preparation, we will compute the second derivative of $\mathcal{Z}[M_4]$. 
From Eq.~\eqref{eq:DMVVmw} we get for the first derivative
\begin{equation}
\begin{split}
\partial_{\bar{y}}  \mathcal{Z}[M_4]\big|_{\bar{y}=1}&=\partial_{\bar{y}} \left(\prod_{n=1}^{\infty} \prod_{\Delta,\bar{\Delta},\ell,\bar{\ell},m_i,w_i}^{'} \frac{1}{(1-p^n q^{\frac{\Delta}{n}+\frac{\delta}{4n}}\bar{q}^{\frac{\bar{\Delta}}{n}+\frac{\bar\delta}{4n}}y^{\ell} \bar{y}^{\bar{\ell}}\zeta_i^{m_i}\xi_i^{w_i})^{c(\Delta,\bar{\Delta},\ell,\bar{\ell},m_i,w_i)}}\right)_{\bar{y}=1}\\
&= \sum_{n,\Delta,\bar{\Delta},\ell,\bar{\ell}}^{'}\frac{\bar{\ell}\, c(\Delta,\bar{\Delta},\ell,\bar{\ell},m_i,w_i)\, p^n q^{\frac{\Delta}{n}+\frac{\delta}{4n}}\bar{q}^{\frac{\bar{\Delta}}{n}+\frac{\bar\delta}{4n}}y^{\ell}  \bar{y}^{\bar{\ell}-1}\zeta_i^{m_i}\xi_i^{w_i}}{(1-p^n q^{\frac{\Delta}{n}+\frac{\delta}{4n}}\bar{q}^{\frac{\bar{\Delta}}{n}+\frac{\bar\delta}{4n}}y^{\ell} \bar{y}^{\bar{\ell}}\zeta_i^{m_i}\xi_i^{w_i})}\mathcal{Z}[M_4]\bigg|_{\bar{y}=1}\\
&=0\,.
\label{eq:partial2YbarCurlyZ}
\end{split}
\end{equation}
Above we have used the 2nd line of \eqref{eq:zero_sums}. Now, we take the second derivative with respect to $\bar{y}$. If the extra derivative hits $\mathcal{Z}$ on the second line of \eqref{eq:partial2YbarCurlyZ} then it will not contribute anything because the first derivative is zero when $\bar{y}=1$. To get a non-vanishing result the derivative needs to act on the sum. When we set the $\bar{y}=1$, then according to the 3rd line of \eqref{eq:zero_sums}  only the term with $\bar{\Delta}=0$ will contribute. With some minor manipulations as done in \cite{Maldacena:1999bp}, we get 
%\footnote{The prime indicates restricting to the quantization condition $\Delta_{total}-\bar{\Delta}_{total} \in \mathbb{Z}$. With $\bar{\Delta}=0$ it implies $\Delta+e_i m_i \in i . \mathbb{Z}$ .}
\begin{eqnarray}
\label{gene}
    \frac{1}{2} \partial^2_{\bar{y}} \mathcal{Z}\big|_{\bar{y}=1}= \sum_{n\geq 1} \sum_{\Delta \geq 0,m_i,w_i}\!\!\!\!\!\!\!'\ \ \sum_{\ell \in \mathbb{Z}} \frac{\hat{c}_2( \Delta, \ell;m_i,w_i) p^n q^{\frac{\Delta}{n}+\frac{\delta}{4n}}\bar{q}^{\frac{\bar\delta}{4n}}y^{\ell}\zeta_i^{m_i}\xi_i^{w_i}}{(1- p^n q^{\frac{\Delta}{n}+\frac{\delta}{4n}}\bar{q}^{\frac{\bar\delta}{4n}}y^{\ell})^2}.
\end{eqnarray}
The new counting function $\hat c_2$ above is defined as
\begin{eqnarray}
    \hat{c}_2 (\Delta,\ell;m_i,w_i):=\frac{1}{2}\sum_{\bar{\ell}}\bar{\ell}^{\,2} c(\Delta,0,\ell,\bar{\ell},m_i,w_i).\label{eq:cHat2Def}
\end{eqnarray}
We will actually take advantage of the fact that the $T^4$ result \eqref{eq:E2_mw_T4} does not depend on $m_i,w_i$, to write $\hat{c}_2 (\Delta,\ell)$ instead of $\hat{c}_2 (\Delta,\ell;m_i,w_i)$ below.

After expanding out \eqref{gene} we get
\begin{eqnarray}
    \frac{1}{2} \partial^2_{\bar{y}} \mathcal{Z}\big|_{\bar{y}=1}=\sum_{s,n,\Delta,\ell,m_i,w_i}\!\!\!\!\!\!\!\!\!' \ \ s\, \big(p^n q^{\frac{\Delta}{n}+\frac{\delta(m_i,w_i)}{4n}}\bar{q}^{\,\frac{\bar\delta(m_i,w_i)}{4n}}y^{\ell}\zeta_i^{\,m_i}\xi_i^{\,w_i}\big)^s \hat{c}_2(\Delta, \ell) \,,\label{eq:curlyZ_s}
\end{eqnarray}
where we have $s,n \geq 1, \Delta \geq 0, \ell \in \mathbb{Z}$. Now, we will also restrict the sum to a particular topological sector as in the definition of the index \eqref{eq:mc_index}:
\begin{eqnarray}\label{eq:partial2yBar_pN}
    \frac{1}{2} \partial^2_{\bar{y}} \mathcal{Z}\big|_{\bar{y}=1}\bigg|_{p^N}=\sum_{n\in\text{Fac}N}\frac{N}{n}\sum_{\Delta,\ell,m_i,w_i}\!\!\!\!\!\!' \ \, \big( q^{\frac{N\Delta}{n^2}+\frac{N\delta(m_i,w_i)}{4n^2}}\bar{q}^{\frac{N\bar{\delta}(m_i,w_i)}{4n^2}}y^{\frac{N\ell}{n}}\zeta_i^{\frac{N\,m_i}{n}}\xi_i^{\frac{N\,w_i}{n}}\big)\, \hat{c}_2(\Delta, \ell)\,. %\\
    %&&=\sum_{m,\ell}  (p^N q^m q^{\frac{\bar{\delta}}{4N}}\bar{q}^{\frac{\bar{\delta}}{4N}}y^{\ell}) \hat{c}(N m- e_i m_i, \ell) 
\end{eqnarray}
Note that defining $m\in\mathbb{Z}$ through
\begin{equation}
    m:=\frac{\Delta+\sum_i m_i w_i}{n},
\end{equation}
in the $n$th sector, we can trivialize the symmetric orbifold constraint \eqref{eq:BPS'}, and write instead
\begin{equation}\label{eq:partial2yBar_pN_simp}
    \frac{1}{2} \partial^2_{\bar{y}} \mathcal{Z}\big|_{\bar{y}=1}\bigg|_{p^N}=\sum_{n\in\text{Fac}N}\!\!\frac{N}{n}\!\!\!\sum_{m,\ell,m_i,w_i} \!\!\! \Big( q^{\frac{N\,m}{n}}(q\bar{q})^{\frac{\bar{\delta}\big(\frac{Nm_i}{n},\frac{Nw_i}{n}\big)}{4N}}y^{\frac{N\ell}{n}}\zeta_i^{\frac{N\,m_i}{n}}\xi_i^{\frac{N\,w_i}{n}}\Big)\ \hat{c}_2(nm-\sum_i m_iw_i, \ell)\,. 
\end{equation}
This suggests that another potentially interesting index can be defined by removing $(q\bar{q})^{\frac{\bar{\delta}}{4N}}$ from the partition function. In the interest of conceptual transparency, however, we stick to the definition \eqref{eq:mc_index}. Focusing on the presumably dominant contribution from $n=N$, we get
\begin{equation}
    \mathcal{E}^{m_i,w_i}_2[\mathrm{sym}^N T^4]\approx\sum_{\Delta,\ell}\,\!' \ \,  q^{\frac{\Delta}{N}}y^{\ell}\, \hat{c}_2(\Delta, \ell)= \sum_{m,\ell}  q^{m-\frac{\sum_i m_iw_i}{N}}y^{\ell}\, \hat{c}_2(Nm-\sum_i m_iw_i, \ell)\,.
\end{equation}
Note that in the $n=N$ sector:
\begin{equation}
    m=\frac{\Delta+\sum_i m_i w_i}{N}.
\end{equation}

To summarize, the degeneracy of the states with 
\begin{equation}
    \begin{split}
        L_0&=m-\frac{\sum_i m_i w_i}{N}+\frac{\delta}{4N}=m+\frac{\bar\delta}{4N},\\
        \bar{L}_0&=\frac{\bar\delta}{4N},\label{eq:L0_E2}
    \end{split}
\end{equation}
and $ \ell=j$, encoded in $\mathcal{E}_2^{m_i,w_i}[\mathrm{sym}^N T^4]$, is (up to the $n<N$ contributions) given by
\begin{equation}
    \hat{c}_2(N m-\sum_i m_iw_i,j).
\end{equation}

Now the goal is to compute $\hat{c}_2(N m-\sum_i m_iw_i,j)$. We first define the single-copy counting function $c_2$ through
\begin{eqnarray}
\label{eq:coeff_E2}
\mathcal{E}^{m_i,w_i}_2[T^4]= \sum_{\Delta,\ell}  c_2(\Delta,\ell;m_i,w_i) q^{\Delta}  y^{\ell}.
\end{eqnarray}
The function $\hat c_2$ is, up to an $\mathcal{O}(1)$ coefficient, equal to the single-copy counting function $c_2$. More explicitly, upon expanding 
\begin{eqnarray}
-  \overline{  \frac{\theta_1(z,\tau)^2}{\eta^6(\tau)}}\Big|_{\bar{q}^0}= \bar{y}+\bar{y}^{-1}-2,
\end{eqnarray}
and evaluating \eqref{eq:cHat2Def} using \eqref{eq:zero_sums}, one finds
\begin{eqnarray}
     \hat{c}_2(N m-\sum_i m_iw_i,j) =c_2(N m-\sum_i m_iw_i,j).
\end{eqnarray}
Here again we have made explicit the independence of the oscillator spectrum from the charge sector $m_i,w_i$. Next, we will calculate $c_2(N m-\sum_i m_iw_i,j)$ using the saddle-point method.

%%%%
\section{Saddle-point analysis and black hole entropy}
\label{sec:saddle_point_T4}

The degeneracy $\tilde{d}(m,N,j,m_i,w_i)={c}_2(N m-\sum_i m_iw_i,j)$ can be calculated via contour integration of $\mathcal{E}^{m_i,w_i}_2[T^4]$, which for simplicity we denote by $\mathcal{H}_2(q,y)$. We have
\begin{equation}
\label{integrand}
\hat{c}_2(N m-\sum_i m_iw_i,j)\simeq \int_0^1 \mathrm{d}\tau \int_0^1
\mathrm{d}z\ e^{-2\pi i \tau  (N m-\sum_i m_iw_i)-2\pi i j z}\
\mathcal{H}_2(q,y).
\end{equation}
\begin{equation}
    \mathcal{H}_2(q,y):= \frac{\theta_1(z,\tau)}{\eta^3(\tau)}\frac{\theta_1(z,\tau)}{\eta^3(\tau)}\nonumber
\end{equation}
Following the analysis in \cite{Sen:2012cj} we notice that the saddle-point value of $|\tau|$
is small, but that of $z$ can be large. We can take  $\mathrm{Re}z$ fixed inside $(0,1)$. Then we expand the Jacobi theta function around these asymptotic values of $\tau$ and $z$ to get
\begin{equation}
\tilde{d}(m,N,j,m_i,w_i)\sim \int_0^1 \mathrm{d}\tau \int_0^1
\mathrm{d}z\ e^{-2\pi i  (N m-\sum_i m_iw_i)\tau-2\pi i j z-2\pi i
z^2\, /\tau+2\pi i z\, /\tau} (\tau)^2.\label{eq:last}
\end{equation}

After extremizing the integrand we find a saddle point at
\begin{equation}
z_0=\frac{1}{2}-\frac{j}{2}\tau_0,\ \ \quad
\tau_0=\frac{i}{2\sqrt{\,  N m-\sum_i m_iw_i-\frac{j^2}{4}}}.
\end{equation}
The maximized exponent at this saddle point has real part
\begin{equation}
S_\text{index}=2\pi\sqrt{ N m-\sum_i m_iw_i\, -\frac{j^2}{4}}.\label{eq:entropyB012}
\end{equation}
This Cardy entropy matches the Bekenstein-Hawking entropy of the bulk black holes \cite{Larsen:1999uk}. \\

What is different compared to the usual uncharged $T^4$ case in \cite{Maldacena:1999bp}? Here, the index calculates the entropy of a black hole that has $(L_0,\bar L_0)=(m,0)+(\frac{\bar\delta}{4},\frac{\bar\delta}{4})$. In particular, for $m_i,w_i\sim\mathcal{O}(N),$ it has $\bar{L}_0=  \bar{\delta}/4 N\sim\mathcal{O}(N)$, instead of zero as in MMS \cite{Maldacena:1999bp}. The nonzero $\bar{L}_0$ is entirely due to the topological charges however. In both cases, the exponential degeneracy is given by the usual oscillator modes. 

\section{Extension to $\mathcal{N}=(2,2)$ AdS$_3$/CFT$_2$}\label{sec:2comma2}

In this main section of the paper, we study the topologically charged states in the HS$_2$ case. Upon the inclusion of charged states, the index will give the right Cardy entropy for a given central charge, resolving the puzzle raised in \cite{ArabiArdehali:2023kar}. In that paper, we looked at the uncharged states (with zero momentum and winding) and found huge bose-fermi cancellations in the index. We got an entropy $S_{\mathrm{index}}= \frac{1}{2} S_{\text{Cardy}}$ from the index for the HS$_2$ case, and similarly found a factor of $1-\frac{1}{k}$ difference in the other $Z_k$ orbifolds. Although we focus on $k=2$ here, we expect that similar conclusions can be drawn for other $k.$

Analogously to the $\mathcal{N}=(4,4)$ case, we will derive the Ramond sector zero-mode algebra, including the topological contributions for the $\mathcal{N}=(2,2)$ theories with momentum and winding charges. Using the Ramond zero-mode algebra, we define a new index, akin to the $\mathcal{N}=(4,4)$ case, that counts the charged states. We will then compute that index in the HS$_2$ sigma model as well as its symmetric orbifold.

Recall that $\mathrm{HS}=T^4/G$, where $T^4=C\times E$, and $G=\mathbb{Z}_k$. The orbifold action is \cite{barth,Datta:2017ert,Eberhardt:2017uup}
\begin{equation}
G\begin{pmatrix} \phi_{12}\\
\phi_{34}
\end{pmatrix}=\begin{pmatrix} \phi_{12}+\frac{2\pi R(1+\tau^{}_C)}{k}\\
e^{2\pi i/k}\,\phi_{34}\\
\end{pmatrix}.\label{eq:GactionZ3}
\end{equation}
Here $\phi_{12}$ and $\phi_{34}$ are two complex scalars which parametrize $C$ and $E$.

In the $k=2$ case, the action of $G=Z_2$ on the supersymmetric $T^4$ sigma model is as follows. On $E$, the $\mathbb{Z}_2$ acts as a reflection (both on the bosonic and fermionic coordinates). On $C$, it acts via target-space translation (it does not act on the fermionic superpartners). 
  
Eberhardt \cite{Eberhardt:2017uup} has argued that these orbifolds reduce the susy from $(4,4)$ to $(2,2)$. In the $T^4$ theory, we have $U(1)^4$ global symmetry, which upon orbifold projection becomes $U(1)^2$. To see the origin of this $U(1)^2$, note that the orbifold acts on $C$ only via translation. Hence, we are left with an $\mathcal{N}=(2,2)$ theory with current algebra given by $U(1)^2.$ We will label the fermionic super partners of these $U(1)^2$ currents by $\bar{Q}^{+}_0,\,\bar{Q}_0^{-}$. These superpartners are nothing but the fermions in the $(2,2)$ theory. One can see this by the action of the supersymmetry generator on the fermions as described below, to conclude $\bar{Q}^{+}_0=\bar{\psi}_0^-$ and $\bar{Q}_0^{-}=\psi_0^-$.

%%%%
\subsection{Modified SUSY algebra in the seed topological sectors}\label{subsec:2comma2_alg}

In this section, we write the mode algebra for the $\mathcal{N}=(2,2)$ theory following \cite{Hori:2003ic}.

Let us start with light-cone coordinates as $x^{\pm}= t\pm \sigma$. The partial derivatives are then
$\partial_{\pm}= \frac{1}{2}(\frac{\partial}{\partial t}\pm\frac{\partial}{\partial \sigma} )$.

The expressions for the anti-holomorphic supercharges are\footnote{Note that unlike in \cite{Hori:2003ic} we use bar to indicate worldsheet (rather than target-space) anti-holomorphic parts. Also, we use $G$ for the supercharge, rather than the supercurrent.}
\begin{eqnarray}
 \bar{G}^{-}_0= \int d\sigma 2 \delta_{i\bar{j}} \partial_{-} \bar{\varphi}^{\bar{j}}\psi^i_{-},\\
 \bar{G}^{+}_0= \int d\sigma 2 \delta_{\bar{i}j} \bar{\psi}^{\bar{i}}_- \partial_{-} {\varphi}^j.
\end{eqnarray}
The $\pm$ label on the supercharges are the eigenvalues of  R-symmetry generator $J_0$. These charges generate the SUSY variation of the fields as:
  \begin{eqnarray}
       \delta \psi_-&&\!\!= -2 i \bar{\epsilon}_+ \partial_- \varphi, \qquad   \delta \bar{\psi}_-= 2 i \epsilon_+ \partial_- \bar{\varphi}\,,\\
       \delta \varphi&&= \epsilon_+ \psi_- -\epsilon_- \psi_+, \quad  \delta \bar{\varphi}=- \bar{\epsilon}_+ \bar{\psi}_-+ \bar{\epsilon}_- \bar{\psi}_+\,.
   \end{eqnarray}
Let us also write the canonical commutation relations for the bosons and fermions:
\begin{eqnarray}
  [\pi^{\nu}(\sigma'),X^{\mu}(\sigma)]=  \eta^{\mu\nu} \delta (\sigma-\sigma'), \quad [\alpha^{\mu}_m,\alpha_n^{\nu}]= m \eta^{\mu\nu} \delta_{m,-n}, \\
  \{\psi^{\mu}_{A}(\sigma),\psi^{\nu}_{B}(\sigma')\}=2 \pi \delta_{AB}\eta^{\mu\nu} \delta (\sigma-\sigma'), \quad   \{\psi_n^{\mu},\psi_m^{\nu}\}= \eta^{\mu\nu}\delta_{n+m,0},
\end{eqnarray}
where $A,B=\pm.$ Here $\psi_n^{\mu}$ is the mode of $\psi^{\mu}_-(\tau,\sigma)$ defined as
 \begin{eqnarray}
       \psi_-^{\mu}= \sum_{n \in \mathbb{Z}} \psi_n^{\mu} e^{-in (t-\sigma)}, \quad \bar{\psi}^\nu_-= \sum_{n \in \mathbb{Z}} \bar{\psi}^\nu_n e^{-in (t-\sigma)}.
   \end{eqnarray}
The mode can be obtained via
\begin{eqnarray}
    \psi_m^{\mu}= \frac{1}{2 \pi}\int d \sigma' \psi_-^{\mu} e^{i m (t-\sigma')},  
\end{eqnarray}
where we have used the identity  $\frac{1}{2 \pi}\sum_n e^{i n (\sigma-\sigma') }= \delta(\sigma-\sigma')$.

We have
\begin{equation}
    \begin{split}
        \bar{Q}^{+}_0&=\bar{\psi}_0^-\,,\\
        \bar{Q}_0^{-}&=\psi_0^-\,,\\
        \lbrace \bar{Q}_0^+,\bar{Q}_0^-\rbrace&= 1,
    \end{split}
\end{equation}
and the action of the R-symmetry charge on $Q$'s is
   \begin{eqnarray}
    [\bar{J}_0, \bar{Q}_0^-]=-\frac{1}{2}\bar{Q}_0^-, \quad  [\bar{J}_0, \bar{Q}_0^+]=+\frac{1}{2}\bar{Q}_0^+.
\end{eqnarray}

The non-trivial anti-commutator of $G$ with $Q$ can be worked out as follows. First, we set the convention for momentum and winding modes. We write the mode expansion for a single compact scalar and then find the momentum and winding modes:   \begin{eqnarray}
   \label{mode_scalar}
  \phi(t,\sigma)&&= \phi_R(t-\sigma)+\phi_L(t+\sigma),\nonumber\\
  \phi_R(t-\sigma)&&= \frac{\phi_0- \hat{\phi}_0}{2}+ \frac{1}{\sqrt{2}} (t-\sigma) p_R +\sum_n\frac{i }{\sqrt{2}} \frac{1}{n} \alpha_n e^{-in (t-s)},\nonumber\\
   \phi_L(t+\sigma)&&= \frac{\phi_0+ \hat{\phi}_0}{2}+ \frac{1}{\sqrt{2}} (t+\sigma) p_L +\sum_n\frac{i }{\sqrt{2}} \frac{1}{n} \tilde{\alpha}_n e^{-in (t+s)}.
\end{eqnarray}
 Above $p^{}_L= \frac{1}{\sqrt{2}} (\frac{m}{R}+ w R)=\frac{u}{\sqrt{2}}$ and  $p^{}_R= \frac{1}{\sqrt{2}} (\frac{m}{R}- w R)=\frac{\bar u}{\sqrt{2}}$.

The commutation relation for the constant modes can be written as
\begin{eqnarray}
    [\phi_0,p_0]=i , \qquad [\hat{\phi}_0,\hat{p}_0]= i.\nonumber
\end{eqnarray}
Here $p_0$ and $\hat{p}_0$ are operators, while their eigenvalues are $$p_0|m,w\rangle= \frac{m}{R} |m,w \rangle,\qquad \hat{p}_0|m,w\rangle= wR\, |m,w \rangle.$$

With this understanding, and defining the complex scalar $\varphi= \frac{1}{\sqrt{2}}(\phi_1+i \phi_2)$, we can evaluate
   \begin{eqnarray}
       \{\bar{G}^-_0,\bar{Q}_0^+\}=\frac{1}{2 \pi}\int d\sigma 2i  \partial_{-} \bar{\varphi}&&= \frac{1}{\sqrt{2}}\frac{i}{\pi}\int d\sigma \partial_-(\phi_1-i \phi_2)\nonumber\\
       &&=\frac{i}{\sqrt{2}} \Big[(\frac{m_1}{R_1}-w_1 R_1)-i  (\frac{m_2}{R_2}-w_2 R_2)\Big]\nonumber\\
      &&\equiv \frac{i}{\sqrt{2}}( \bar u_1- i \bar u_2)\,.\label{eq:anti_comm_1}
        \end{eqnarray}
The second line of the above equation follows directly from the mode expansion of the compact scalar \eqref{mode_scalar}. Similarly, we can work out the other anti-commutator as
\begin{eqnarray}
       \{ \bar{G}^{+}_0,\bar{Q}^{-}_0\}=\frac{-2i}{2 \pi}\int d\sigma   \partial_{-} \varphi= &&=-\frac{i}{\sqrt{2}} \Big[(\frac{m_1}{R_1}-w_1 R_1)+i  (\frac{m_2}{R_2}-w_2 R_2)\Big]\nonumber\\
       &&\equiv -\frac{i}{\sqrt{2}}( \bar u_1+ i \bar u_2)\,.\label{eq:anti_comm_2}
\end{eqnarray}

In equations \eqref{eq:anti_comm_1}, \eqref{eq:anti_comm_2}, we have treated the scalar $\varphi$ as if it stands for the un-orbifolded torus scalar. To adapt the results to the $\mathbb{Z}_2$ orbifold setting, we impose that $w_i$ are integer-valued in the untwisted sector, and half-odd-integers in the twisted sector. Later we also adapt the results to the symmetric orbifold setting via the DMVV formula.

In summary, the algebra of interest can be written as
 \begin{eqnarray}
 \label{eq:zero_mode_susy_hs}
    &&\lbrace \bar{G}_0^+,\bar{G}_0^-\rbrace= 2 \bar{L}_0, \quad \{\bar{G}_0^-,\bar{Q}^+_0\}= \frac{i}{\sqrt{2}}(\bar u_1- i \bar u_2)\,,\nonumber\\
    &&\lbrace \bar{Q}_0^+,\bar{Q}_0^-\rbrace= 1, \quad\quad \{\bar{G}_0^+,\bar{Q}_0^-\}= -\frac{i}{\sqrt{2}}(\bar u_1+ i \bar u_2)\,,\nonumber\\
   &&  [\bar{J}_0,\bar{G}_0^\pm]=\pm\frac{1}{2} \bar{G}_0^\pm,\nonumber\\
    &&  [\bar{J}_0,\bar{Q}_0^\pm]=\pm\frac{1}{2} \bar{Q}_0^\pm.\label{eq:2,2alg}
 \end{eqnarray}

%%%%
\subsection{Shortening condition and BPS states}

Now let us identify the creation operators $d_i^{\dagger}$ as $\bar{G}_0^{+},\bar{Q}_0^+$, and the annihilation operators $d_i$ as $\bar{G}_0^{-},\bar{Q}_0^-$. We expect to find
\begin{equation}
    \{d_i,d_j\}=0,\quad \{d^\dagger_i,d^\dagger_j\}=0,\quad \{d_i,d^\dagger_j\}=M_{ij},
\end{equation}
with a $2\times 2$ Hermitian matrix $M_{ij}.$ 
The matrix $M_{ij}$ is found through explicit calculation as
 \begin{eqnarray}
   \left(
\begin{array}{cccc}
 2 \bar{L}_0 & -\frac{i}{\sqrt{2}} (\bar u_1+i \bar u_2) \\
\frac{i}{\sqrt{2}} (\bar u_1-i \bar u_2) & 1  \\
\end{array}
\right)   .
 \end{eqnarray}
The (anti-chiral) topological charges $\bar u_i=( \frac{m_i}{R_i}- w_i R_i) $ are real. Recall that $R_{1,2}$ are the un-orbifolded radii, while $w_i$ are integers in the untwisted sector and half-odd-integers in the twisted sector.  Diagonalizing this $M_{ij}$ matrix, the eigenvalues are found to be
\begin{eqnarray}
  \left\{-\frac{1}{2} \sqrt{4 \bar{L}_0^2-4 \bar{L}_0+2
   \bar u_1^2+2 \bar u_2^2+1}+\bar{L}_0+\frac{1}{2},\frac{1}{2}
   \left(\sqrt{4 \bar{L}_0^2-4 \bar{L}_0+2 \bar u_1^2+2
   \bar u_2^2+1}+2 \bar{L}_0+1\right)\right\}.\nonumber\\
\end{eqnarray}
When $\bar{L}_0= \frac{1}{4}(\bar u_1^2+\bar u_2^2)$,  one of the eigenvalues becomes zero. We thus get a "shortening'' condition. The BPS bound is hence
\begin{equation}
    \bar{L}_0\ge \frac{1}{4}(\bar u_1^2+\bar u_2^2)\,,
\end{equation}
with  $\bar u_i=(\frac{m_i}{R_{i}}- w_i R_{i})$.

It is straightforward to check that the fermionic creation operator becoming null on the BPS states is
\begin{equation}
    \boxed{\bar{\mathbbmtt{Q}}_{\,m_i,w_i}^\dagger:=\bar G^+_0+i\big(\frac{\bar u_1+i\bar u_2}{\sqrt{2}}\big)\bar Q^+_0\,.}\label{eq:deformed_Q}
\end{equation}

%%%%
\subsection*{From the seed to the symmetric orbifold}

Similarly to Section~\ref{sec:shortening_T4}, the BPS bound in the $N$-fold symmetric orbifold is $N$ times smaller than that in the seed:
\begin{equation}
    \bar{L}_0\ge\frac{1}{4N}(\bar u_1^2+\bar u_2^2).
\end{equation}

We are interested in counting the states saturating the above bound in $\mathrm{sym}^N(\mathrm{HS}_2)$, in an arbitrary momentum and winding sector labeled by $m_{1,2},w_{1,2}.$

%%%%
\subsection{Encoding the BPS states into partition functions}

Considerations similar to those in Section~\ref{sec:pn_fns_T4} force us to settle for the following analog of the first helicity-trace index used in \cite{ArabiArdehali:2023kar}:
\begin{equation}
    \mathcal{E}^{m_i,w_i}_1(q,y)=\mathrm{Tr}_{m_i,w_i}\,(-1)^{2J_0-2\bar J_0}\,2\bar J_0\,q^{L_0-\frac{\delta(m_i,w_i)}{4N}}\,y^{2J_0}.\label{eq:mc_index_E1}
\end{equation}
We emphasize that we have chosen to define of the quantum numbers $m_{1,2},\, w_{1,2}$ with respect to $C\subset T^4$ (and extend to $\mathrm{sym}^N\mathrm{HS}_2$, first via the $\mathbb{Z}_2$ orbifold \eqref{eq:orbTrZ2}, and then via DMVV). In other words, our $m_i,\, w_i$ are not defined intrinsically with respect to HS$_2$ (or $\mathrm{sym}^N\mathrm{HS}_2$).

Explicit evaluation of $\mathcal{E}_1$ starts again with the partition function, which is already computed for the $\mathrm{HS}_2$ case in \cite{ArabiArdehali:2023kar}:
\begin{equation}
\begin{split}
Z[\mathrm{HS}_2]&=\frac{1}{2}\Theta^{T^4}\left|\frac{\theta_1(z,\tau)}{\eta^3}\right|^4
+2\left|\frac{\theta_2(z,\tau)}{\theta_2(\tau)}\right|^2\cdot
\Theta^{T^2}_{\text{w/\ }G\text{-insertion}}\cdot
\left|\frac{\theta_1(z,\tau)}{\eta^3}\right|^2\\
&\ \ +2\left|\frac{\theta_4(z,\tau)}{\theta_4(\tau)}\right|^2\cdot
\Theta^{T^2}_{\text{w/\ }1/2\text{-shifted lattice}}\cdot
\left|\frac{\theta_1(z,\tau)}{\eta^3}\right|^2\\
&\ \ +2\left|\frac{\theta_3(z,\tau)}{\theta_3(\tau)}\right|^2\cdot
\Theta^{T^2}_{\text{w/\ }G\text{-insertion, }1/2\text{-shifted
lattice}}\cdot \left|\frac{\theta_1(z,\tau)}{\eta^3}\right|^2.\label{eq:Zhs2_main}
\end{split}
\end{equation}
There are four terms in the partition function, corresponding to the untwisted (first line) and twisted sector (second and third lines). We know the asymptotic properties of these Jacobi theta functions (see \emph{e.g.}~Appendix~C of \cite{ArabiArdehali:2023kar}). It turns out that for the BTZ saddle ($c=1,d=0$---here $c,d$ label the $SL(2,\mathbb{Z})$ family of saddles), the term in the second line of \eqref{eq:Zhs2_main} (twisted sector but no $G$ insertion) will give the maximum growth in the Cardy limit $\tau\to0$.\footnote{We have analyzed all other terms and they all are subleading compared to the third term which gives the leading Bekenstein-Hawking/Cardy entropy.}  Hence, we will focus on this term:
\begin{eqnarray}
   Z[\mathrm{HS}_2]&& \supset2\left|\frac{\theta_4(z,\tau)}{\theta_4(\tau)}\right|^2\cdot
\left|\frac{\theta_1(z,\tau)}{\eta^3}\right|^2 \Bigg(\Theta^{T^2}_{w/\text{ 1/2-shifted lattice}}\Bigg).
\end{eqnarray}
Let's write the shifted lattice explicitly (details can be found in Appendix \ref{app:HS2}):
\begin{equation}
    \Theta^{T^2}_{\text{w{\tiny/}}\,\frac{1}{2}\text{-shifted lattice}}=\sum_{\substack{w_i\in\,\mathbb{Z}+\frac{1}{2}\\ m_i \in\,\mathbb{Z}}}q^{\frac{1}{4}\sum_i^2(\frac{m_i}{R_i}+{w_iR_i}{})^2}\bar{q}^{\frac{1}{4}\sum_i^2(\frac{m_i}{R_i}-{w_iR_i}{})^2}.\label{eq:ThetaT2shited}
\end{equation}
The radii $R_i$ in the above expression are those of the original $T^4$.

\begin{comment}
From the BPS bound we have $\bar{L}_0=\frac{1}{4}(u_1^2+u_2^2)=\frac{1}{4}( \frac{a_1}{R_{hs}}- b_1 R_{hs})^2+\frac{1}{4}( \frac{a_2}{R_{hs}}- b_2 R_{hs})^2 $. We can write this interms of $T^4$ radii using the fact that $R_{hs}= \frac{R_{T^4}}{2}\equiv \frac{R_i}{2} $. Then the BPS bound will become 
\begin{eqnarray}
    \bar{L}_0= \frac{1}{4} \sum_i^2 \Big(\frac{2 a_i}{ R_i}- \frac{b_i R_i}{2 } \Big)
\end{eqnarray}
Now we identify $b_i= m_i, 2 a_i= e_i$, Then the charges saturating the BPS bound ($a_i ,b_i$) will be present in the shifted lattice. It means the index defined above will pick out the terms in the shifted lattice having $e_i \in even$ and $m_i \in odd$. 
\end{comment}

Committing to a specific topological sector we get
\begin{eqnarray}
    Z^{m_i,w_i}(q,\bar{q},y,\bar y)&&= 2\left|\frac{\theta_4(z,\tau)}{\theta_4(\tau)}\right|^2\cdot
\left|\frac{\theta_1(z,\tau)}{\eta^3}\right|^2 q^{\frac{1}{4}\sum_{i=1}^2(\frac{m_i}{R_i}+{w_iR_i}{})^2}\bar q^{\frac{1}{4}\sum_{i=1}^2(\frac{m_i}{R_i}-{w_iR_i}{})^2}\nonumber\\
&&\equiv 2\left|\frac{\theta_4(z,\tau)}{\theta_4(\tau)}\right|^2\cdot
\left|\frac{\theta_1(z,\tau)}{\eta^3}\right|^2 q^{\frac{\delta}{4}}\bar q^{\frac{\bar\delta}{4}}.\label{eq:Z_HS2_miwi}
\end{eqnarray}
Here we have defined $\delta=\sum_{i=1}^2(\frac{m_i}{R_i}+{w_iR_i}{})^2$, and analogously for $\bar\delta$. 

Comparison of the partition function \eqref{eq:Z_HS2_miwi} and the index \eqref{eq:mc_index_E1} shows that the index can be obtained by setting $\bar q=1$ in $Z^{m_i,w_i}$, removing the $q^{\delta/4}$, and finally taking $\partial_{\bar y}|_{y=1}$ to implement the insertion of $2\bar{J}_0$. The result is
\begin{equation}
    \boxed{\mathcal{E}^{m_i,w_i}_1[\mathrm{HS}_2]\supset-2i\,\frac{\theta_1(z,\tau)}{\eta^3}\frac{\theta_4(z,\tau)}{\theta_4(\tau)}=2\big(y^{-\frac{1}{2}}-y^{\frac{1}{2}}\big)+\mathcal{O}\big(q^{1/2}\big),}\label{eq:E1_mw_HS}
\end{equation}
where we have emphasized by $\supset$ that only one term in \eqref{eq:Zhs2_main} is considered. Note that, as in the $T^4$ case, the contribution of our interest to $\mathcal{E}_1^{m_i,w_i}[\mathrm{HS}_2]$ is the same irrespective of the topological sector. Again, that is because it simply counts the oscillator excitations on top of a topological ``vacuum'' $|m_i,w_i\rangle$.

Next, we consider the symmetric orbifold. 

%%%%
\subsection*{From the seed to the symmetric orbifold via DMVV}
\label{symorbhs2}

We write out the above partition function as
\begin{eqnarray}
\label{coeffhs2}
Z^{m_i,w_i}(q,\bar{q},y,\bar y)= \sum_{\Delta,\bar{\Delta},\ell,\bar{\ell}}  c(\Delta,\bar{\Delta},\ell,\bar{\ell}) q^{\Delta} \bar{q}^{\bar{\Delta}} y^{\ell} \bar{y}^{\bar{\ell}} q^{\delta/4}\bar q^{\bar\delta/4}   
\end{eqnarray}

We will calculate the 1st helicity-trace index of the symmetric orbifold theory with target space $\mathrm{sym}^N \mathrm{HS}_2$.  We start with the generating function of the symmetric orbifold theories given by the DMVV formula as
\begin{eqnarray}
    \mathcal{Z}[M_4]=\prod_{n=1}^{\infty} \prod_{\Delta,\bar{\Delta},\ell,\bar{\ell},m_i,w_i}^{'} \frac{1}{(1-p^n q^{\frac{\Delta}{n}+\frac{\delta}{4n}}\bar{q}^{\frac{\bar{\Delta}}{n}+\frac{\bar\delta}{4n}}y^{\ell} \bar{y}^{\bar{\ell}}\zeta_i^{m_i}\xi_i^{w_i})^{c(\Delta,\bar{\Delta},\ell,\bar{\ell})}}.\label{eq:DMVVoghs2}
\end{eqnarray}
The prime on the product means that $\Delta,\bar{\Delta}$ are restricted such that $\frac{\Delta-\bar\Delta+\sum_i m_i w_i}{n}\in\mathbb{Z}$. Again in the BPS sectors where $\bar\Delta=0$ we have
\begin{equation}
    \frac{\Delta+\sum_i m_i w_i}{n}\in\mathbb{Z}.\label{eq:BPS'2}
\end{equation}
We also keep in mind that the states of our interest have $w_i\in\,\mathbb{Z}+\frac{1}{2}.$

The coefficients $c(\Delta,\bar{\Delta},\ell,\bar{\ell})$ represent the coefficients of the seed partition function \eqref{coeffhs2}.  From Eq.~\eqref{eq:DMVVoghs2} the derivative of $\mathcal{Z}[M_4]$ with respect to $\bar{y}$ gives
\begin{equation}
\begin{split}
\partial_{\bar{y}}  \mathcal{Z}[M_4]\big|_{\bar{y}=1}&=\partial_{\bar{y}} \left(\prod_{n=1}^{\infty} \prod_{\Delta,\bar{\Delta},\ell,\bar{\ell},m_i,w_i}^{'} \frac{1}{(1-p^n q^{\frac{\Delta}{n}+\frac{\delta}{4n}}\bar{q}^{\frac{\bar{\Delta}}{n}+\frac{\bar\delta}{4n}}y^{\ell} \bar{y}^{\bar{\ell}}\zeta_i^{m_i}\xi_i^{w_i})^{c(\Delta,\bar{\Delta},\ell,\bar{\ell})}}\right)_{\bar{y}=1}\\
&=\sum_{n,\Delta,\bar{\Delta},\ell,\bar{\ell},m_i,w_i}^{'}\frac{\bar{\ell}\, c(\Delta,\bar{\Delta},\ell,\bar{\ell})\, p^n q^{\frac{\Delta}{n}+\frac{\delta}{4n}}\bar{q}^{\frac{\bar{\Delta}}{n}+\frac{\bar\delta}{4n}}y^{\ell}  \bar{y}^{\bar{\ell}-1}\zeta_i^{m_i}\xi_i^{w_i}}{(1-p^n q^{\frac{\Delta}{n}+\frac{\delta}{4n}}\bar{q}^{\frac{\bar{\Delta}}{n}+\frac{\bar\delta}{4n}}y^{\ell} \bar{y}^{\bar{\ell}}\zeta_i^{m_i}\xi_i^{w_i})}\mathcal{Z}[M_4]\bigg|_{\bar{y}=1}\\
&=\sum_{n \in \mathbb{N}, \Delta\geq 0,\ell,m_i,w_i}' \frac{\hat{c}_1(\Delta,\ell) p^n q^{\frac{\Delta}{n}+\frac{\delta}{4n}}\bar{q}^{\frac{\bar\delta}{4n}}y^{\ell}\zeta_i^{m_i}\xi_i^{w_i}}{1- p^n q^{\frac{\Delta}{n}+\frac{\delta}{4n}}\bar{q}^{\frac{\bar\delta}{4n}}y^{\ell}\zeta_i^{m_i}\xi_i^{w_i}},
\label{eq:partialYbarCurlyZ}
\end{split}
\end{equation}
In the third line we have used $\mathcal{Z}[M_4]\big|_{\bar{y}=1}=1,$ which follows from having the fermion zero mode. 
We have also
used
\begin{equation}
\sum_{\bar{\ell}}\bar{\ell} c(\Delta,\bar{\Delta},\ell,\bar{\ell})=0, \quad \text{for} \quad \bar{\Delta}>0,
\end{equation}
 and defined
\begin{eqnarray}
    \hat{c}_1 (\Delta,\ell):=\sum_{\bar{\ell}}\bar{\ell} c(\Delta,0,\ell,\bar{\ell}).\label{eq:cHat1Def}
\end{eqnarray}
The first identity can be explicitly verified at low values of $\Delta$ in the twisted sector by working out the expansion. It also follows from the SUSY algebra \eqref{eq:zero_mode_susy_hs} by using the fact that  $\bar{G}_0^+,\bar{Q}_0^+$ actually commute with $L_0,\bar{L}_0, J_0$. 
%$\sum_{N=1}^\infty p^N \mathcal{E}_1[\mathrm{sym}^N M_4]=$
After expanding out Eq.~\eqref{eq:partialYbarCurlyZ}, we get
\begin{eqnarray}
 \partial_{\bar{y}}  \mathcal{Z}[M_4]\big|_{\bar{y}=1}=\sum_{s,n \in \mathbb{N}, \Delta\geq 0,\ell,m_i,w_i}' \hat{c}_1(\Delta,\ell)( p^n q^{\frac{\Delta}{n}+\frac{\delta}{4n}}\bar{q}^{\frac{\bar\delta}{4n}}y^{\ell}\zeta_i^{m_i}\xi_i^{w_i})^s .\label{eq:modifiedDMVV}
\end{eqnarray}
Here we have $s,n \geq 1, \Delta \geq 0, \ell \in \mathbb{Z}$. We now restrict to a single topological sector as prescribed in \eqref{eq:mc_index_E1}, and also take $s=1$ and $n=N$:
\begin{eqnarray}
   \partial_{\bar{y}}  \mathcal{Z}[M_4]\big|_{\bar{y}=1}\bigg|_{p^N\zeta_i^{m_i}\xi_i^{w_i}}\approx&& \sum_{\Delta,\ell}'  \ ( q^{\frac{\Delta}{N}+\frac{\delta}{4N}}\bar{q}^{\frac{\bar{\delta}}{4N}}y^{\ell}) \ \hat{c}_1(\Delta, \ell) \nonumber\\
    &&=\sum_{m,\ell}  \ \big( q^{m-\frac{\sum_i m_iw_i}{N}} q^{\frac{{\delta}}{4N}}\bar{q}^{\frac{\bar{\delta}}{4N}}y^{\ell}\big)\  \hat{c}_1(Nm-\!\sum_i m_iw_i, \ell)\,. 
\end{eqnarray}
In the second line above, we have taken $\Delta= Nm-\!\sum_i m_iw_i,\ m \in \mathbb{Z}$, to trivialize \eqref{eq:BPS'2}. 
 %The shift in $L_0^{shift}=\bar{L}_0^{shift}= \frac{\bar{\delta}}{4 N}$.
 
To summarize, the degeneracy of the states with 
\begin{equation}
    \begin{split}
        L_0&=m-\frac{\sum_i m_i w_i}{N}+\frac{\delta}{4N}=m+\frac{\bar\delta}{4N},\\
        \bar{L}_0&=\frac{\bar\delta}{4N},\label{eq:L0_E1}
    \end{split}
\end{equation}
and $\ell=j$, encoded in $\mathcal{E}_1^{m_i,w_i}[\mathrm{sym}^N \mathrm{HS}_2]$, is (up to the $n<N$ contributions, and up to the terms dropped in \eqref{eq:E1_mw_HS}) given by
\begin{equation}
    \hat{c}_1(N m-\sum_i m_iw_i,j),
\end{equation}
assuming $w_i\in\mathbb{Z}+\frac{1}{2}$.

Now the goal is to compute $\hat{c}_1(N m-\sum_i m_iw_i,j)$. We first define $c_1$ through
\begin{eqnarray}
\label{eq:coeff_E1}
\mathcal{E}^{m_i,w_i}_1[\mathrm{HS}_2]= \sum_{\Delta,\ell}  c_1(\Delta,\ell;m_i,w_i) q^{\Delta}  y^{\ell}.
\end{eqnarray}
Similarly to Section~\ref{sec:pn_fns_T4}, the function $\hat{c}_1$ is, up to an $\mathcal{O}(1)$ coefficient, equal to the single copy counting function $c_1(N m-\sum_i m_iw_i,j)$.
\begin{comment}
More explicitly \arash{...}
\begin{eqnarray}
    \hat{c}_1(\Delta,j) =\# c_1(\Delta,j).
\end{eqnarray}
\end{comment}

Next, we will calculate $\hat{c}_1\sim c$ by the saddle point method.

%%%%
\subsection{Saddle-point analysis and black hole entropy }

The degeneracy function $\tilde{d}(m,N,j,m_i,w_i)=\hat{c}_1(N m-\sum_i m_iw_i,j),$ will be calculated via contour integration of $\mathcal{E}^{m_i,w_i}_1[\mathrm{HS}_2]$, which for simplicity we denote by $\tilde{\mathcal{H}}_1(q,y)$. We thus have
\begin{equation}
\label{integrand}
\tilde{d}(m,N,j,m_i,w_i)\overset{w_i\in\,\mathbb{Z}+\frac{1}{2}}{\approx} \int_0^1 \mathrm{d}\tau \int_0^1
\mathrm{d}z\ e^{-2\pi i  \tau (N m-\sum_i m_iw_i)-2\pi i j z}\
\tilde{\mathcal{H}}_1(q,y),
\end{equation}
\begin{equation}
    \tilde{\mathcal{H}}_1(q,y)=\frac{\theta_4(z,\tau)}{\theta_4(\tau)}\frac{\theta_1(z,\tau)}{\eta^3(\tau)},\nonumber
\end{equation}
up to an irrelevant overall numerical factor.

Following the analysis in \cite{Sen:2012cj} we notice that the saddle-point value of $|\tau|$
is small but that of $z$ can be large. We can take  $\mathrm{Re}z$ fixed inside $(0,1)$. Then we expand the Jacobi theta function around these asymptotic values of $\tau$ and $z$ we have 
\begin{equation}
\tilde{d}(m,N,j,m_i,w_i)\overset{w_i\in\,\mathbb{Z}+\frac{1}{2}}{\approx} \int_0^1 \mathrm{d}\tau \int_0^1
\mathrm{d}z\ e^{-2\pi i  (N m-\sum_i m_iw_i)\tau-2\pi i j z-2\pi i
z^2\, /\tau+2\pi i z\, /\tau} \ \tau.\label{eq:last}
\end{equation}
Extremizing the integrand we find a saddle point at
$z_0=\frac{1}{2}-\frac{j}{2}\tau_0\,,\ 
\tau_0=\frac{i}{2\sqrt{\, N m-\sum_i m_iw_i-\frac{j^2}{4}}}\,.$ At the saddle the maximized exponent has real part
\begin{equation}
S_\text{index}=2\pi\sqrt{N m-\sum_i m_iw_i\, -\frac{j^2}{4}}\, .\label{eq:entropyB012}
\end{equation}
This is the Cardy entropy for the symmetric orbifold theory with HS$_2$ seed.

%%%%%
\section{Discussion}\label{sec:discussion}

In this work we extended the modified supersymmetric indices of \cite{Maldacena:1999bp} and \cite{ArabiArdehali:2023kar} to the nonzero momentum and winding sectors. In the $\mathcal{N}=(4,4)$ $T^4$ case we reproduced the results of \cite{Larsen:1999uk}, while in the $\mathcal{N}=(2,2)$ $\mathrm{HS}_2$ case we resolved the black hole
microstate counting mismatch of \cite{ArabiArdehali:2023kar} by going to the charged sectors with $w_i\in\mathbb{Z}+\frac{1}{2}$. The resolution is due to the counting function in \eqref{eq:E1_mw_HS}, arising from the 3rd term on the RHS of \eqref{eq:Zhs2_main}, taking over for $w_i\in\mathbb{Z}+\frac{1}{2}$ (instead of the problematic \eqref{eq:HS2wjf}, which takes over for $w_i\in\mathbb{Z}$, and which arose in \cite{ArabiArdehali:2023kar} from the 2nd term of \eqref{eq:Zhs2_main}).

The corresponding charged black holes are extremal and BPS, but have $\bar{L}_0>0.$ This makes them unusual from the point of view of supersymmetric indices (which usually count states with $\bar{L}_0=0$). We thus had to slightly tweak the usual definition of helicity-trace indices to encode the charged states into protected quantities. In particular, the corresponding supercharge is deformed in the charged sectors as in \eqref{eq:deformed_Q}. The tweak is, however, analogous to the one needed in the context of BPS state counting on the Coulomb branch of 4d $\mathcal{N}=2$ theories, where the SUSY algebra also receives central extension, somewhat similarly to the algebras in Sections~\ref{sec:T4_alg} and \ref{sec:2comma2}. We leave clarification of the relation with the $\mathcal{N}=2$ gauge theory context (as in \emph{e.g.}~\cite{Gaiotto:2010be}), through dimensional reduction, for future work.

Despite the satisfactory resolution to the mismatch of \cite{ArabiArdehali:2023kar} that we have found in the charged sectors with $w_i\in\mathbb{Z}+\frac{1}{2}$, the mismatch in the uncharged sector, and more generally in sectors with $w_i\in\mathbb{Z}$, still calls for a bulk explanation. It would be interesting if there is a connection with the $Q_5$ constraints reviewed in Section~4 of \cite{ArabiArdehali:2023kar}. 

The boundary explanation for the mismatch of \cite{ArabiArdehali:2023kar} appears to be bose-fermi cancellations (\emph{cf.}~\cite{Vafa:1997gr}): the 1st helicity-trace index does not see the BPS states in the first term on the RHS of \eqref{eq:Zhs2_main}, due to the extra fermionic zero-mode which that term effectively has (as it descends from $T^4$). It may be that protected variants of the 1st helicity-trace index of \cite{ArabiArdehali:2023kar}, with additional phase insertions inside the trace, corresponding to various symmetry-twisted boundary conditions, can obstruct the bose-fermi cancellations underlying the mismatch and resolve it in the  sectors with $w_i\in\mathbb{Z}$. We leave investigation of that possibility to the future.

Finally, it is tempting to speculate that the erratic dependence of the microscopic degeneracy on the momentum and winding charges, is another face of the erratic BPS spectrum in the topologically trivial sector of $\mathrm{sym}^N(\mathrm{HS}_2)$ as a function of energy, encountered in Appendix~D of \cite{ArabiArdehali:2023kar}. These might signify BPS chaos \cite{Chen:2024oqv}. It would in any case be interesting to study $\mathrm{sym}^N(\mathrm{HS}_2)$ in connection with \cite{Chen:2024oqv}.

\begin{acknowledgments}

We are indebted to O.~Aharony, Y.~Choi, J.~Distler, M.~Gaberdiel, M.~Heydeman, J.~Maldacena, S.~Murthy, M.~Ro\v{c}ek, Y.~Sanghavi, S.~Seifnashri, S.~Shao, C.~Vafa, and S.~Vandoren for helpful discussions and correspondences. This work was supported in part by the NSF grant PHY-2210533 and the Simons Foundation grants 397411 (Simons Collaboration on the Nonperturbative Bootstrap) and 681267 (Simons Investigator Award). The work of HK is supported by NSF grant PHY 2210533.

\end{acknowledgments}

%%%%%%%%%%%%%%%%%%%%%%%%%%%%%%%%%%%%%%%%%%

%%%%%%%%%%%%%%%%%%%%%%%%%%%%%%%%%%%%%%%%%%%%%
\appendix

\section{DMVV formula with the inclusion of topological charges}
\label{app:dmvv}

In this appendix, we incorporate the topologically charged states carrying target-space momenta and windings, into the DMVV formula.

Define the grand-canonical free energy $\mathcal{F}$ via
\begin{equation}
    \mathcal{Z}[M_4]:=\sum_{N\ge0}p^N Z[\mathrm{sym}^N M_4]\equiv e^{\mathcal{F}[M_4]}.
\end{equation}
One can write $\mathcal{F}(q,\bar{q},y,\bar{y})$ in terms of Hecke operators (generalizing  the construction of \cite{Dijkgraaf:1996xw} to include anti-holomorphic fugacities $\bar q,\bar y$, as well as the topological sectors)
\begin{eqnarray}
 \mathcal{F}(q,\bar{q},y,\bar{y})= \sum_{N>0} p^N T_N \chi (M;q,y,\bar{q},\bar{y})   .
\end{eqnarray}
The action of Hecke operators $T_N$ on the seed partition function can be written as
\begin{eqnarray}
    T_N \chi(q,y,\bar{q},\bar{y})= \sum_{\substack{ad= N\\
     b \,mod \, d}} \frac{1}{N} \chi(\frac{a \tau +b}{d},\frac{a \bar{\tau} +b}{d}, a z, a \bar{z}).
    \end{eqnarray}
The seed partition function $\chi(q,\bar{q},y,\bar{y})$ has Fourier expansion
\begin{eqnarray}
 \chi(q,\bar{q},y,\bar{y})= \sum_{m_i,w_i} q^{\frac{\delta}{4}} \bar{q}^{\frac{\bar{\delta}}{4}} \sum_{\Delta,\bar{\Delta},\ell,\bar{\ell}} c(\Delta,\bar{\Delta},\ell,\bar{\ell}) q^{\Delta} \bar{q}^{\bar{\Delta}} y^{\ell} \bar{y}^{\bar{\ell}} .  
\end{eqnarray}
Here $\delta= \sum_i (\frac{m_i}{R_i}+w_i R_i)^2$ and $\bar{\delta}=\sum_i (\frac{m_i}{R_i}-w_i R_i)^2$. The action of Hecke operators on the seed partition function will have the Fourier expansion
\begin{eqnarray}
    T_N \chi=\sum_{\substack{ad= N\\
     b \,mod \, d}} \frac{1}{a d} \sum_{\Delta,\bar{\Delta},\ell,\bar{\ell},m_i,w_i} c(\Delta,\bar{\Delta},\ell,\bar{\ell}) q^{\frac{a}{d}\Delta+\frac{a}{d}\frac{\delta}{4}} \bar{q}^{\frac{a}{d}\bar{\Delta}+\frac{a}{d}\frac{\bar{\delta}}{4}} y^{a\ell} \bar{y}^{ a\bar{\ell}}  e^{2 \pi i \frac{b}{d}(\Delta- \bar{\Delta}+m_i w_i)}.
\end{eqnarray}
Now, we are going to make the following choice
\begin{eqnarray}
\label{restrcition}
    (\Delta- \bar{\Delta}+ m_i w_i) \in d. \mathbb{Z}.
\end{eqnarray}
With this choice, the sum over $b$ can be trivially done and it cancels with the factor of $d$ in the denominator. We are now left with
\begin{eqnarray}
   T_N \chi=\sum_{ad=N} \frac{1}{a}   \sum_{\Delta,\bar{\Delta},\ell,\bar{\ell},m_i,w_i} c(\Delta,\bar{\Delta},\ell,\bar{\ell})q^{\frac{a}{d}\Delta+\frac{a}{d}\frac{\delta}{4}} \bar{q}^{\frac{a}{d}\bar{\Delta}+\frac{a}{d}\frac{\bar{\delta}}{4}} y^{a\ell} \bar{y}^{ a\bar{\ell}}.
\end{eqnarray}
Hence, the free energy can be written as
\begin{eqnarray}
    \mathcal{F}&&= \sum_{d>0} \sum_{a>0}\frac{1}{a} \sum_{\Delta,\bar{\Delta},\ell,\bar{\ell},m_i,w_i} c(\Delta,\bar{\Delta},\ell,\bar{\ell})  p^{ad}q^{\frac{a}{d}\Delta+\frac{a}{d}\frac{\delta}{4}} \bar{q}^{\frac{a}{d}\bar{\Delta}+\frac{a}{d}\frac{\bar{\delta}}{4}} y^{a\ell} \bar{y}^{ a\bar{\ell}}\nonumber\\
   && =-\sum_{d>0}\sum_{\Delta,\bar{\Delta},\ell,\bar{\ell},m_i,w_i} c(\Delta,\bar{\Delta},\ell,\bar{\ell}) \, \mathrm{log} (1- p^d q^{\frac{\Delta}{d}+\frac{\delta}{4d}}\bar{q}^{\frac{\bar{\Delta}}{d}+\frac{\delta}{4d}}y^{\ell} \bar{y}^{\bar{\ell}}).\label{eq:app_F}
\end{eqnarray}
Note that the constraint $ad=N$ is already imposed as $p^{ad}$.

Incorporating $\zeta_i,\xi_i$ is straightforward, and the generating function can be written as
\begin{equation}
    \mathcal{Z}(q,\bar{q},y,\bar{y},\zeta_i,\xi_i)\equiv e^{\mathcal{F}}=  \prod_{n=1}^{\infty} \prod_{\Delta,\bar{\Delta},\ell,\bar{\ell},m_i,w_i}^{'} \frac{1}{(1-p^n q^{\frac{\Delta}{n}+\frac{\delta}{4n}}\bar{q}^{\frac{\bar{\Delta}}{n}+\frac{\delta}{4n}}y^{\ell} \bar{y}^{\bar{\ell}}\zeta_i^{m_i}\xi_i^{w_i})^{c(\Delta,\bar{\Delta},\ell,\bar{\ell},m_i,w_i)}}.
\end{equation}
The prime in product refers to the restriction that we have in \eqref{restrcition}, with $d$ replaced by $n$.

\section{Topological sector contributions in the $\mathrm{HS}_2$ partition function}\label{app:HS2}

In \cite{ArabiArdehali:2023kar} partition functions were computed for $\mathcal{N}=(2,2)$ orbifolds $\mathrm{HS}_G:=T^4/G$ \cite{Eberhardt:2017uup}, with $G=\mathbb{Z}_k$ for various specific choices of $k$. The $T^4$ can be written in the form of a two complex-dimensional manifold as $T^4=C\times E$. The $\mathbb{Z}_k$ action is then 
\begin{equation}
G\begin{pmatrix} \phi_{12}\\
\phi_{34}
\end{pmatrix}=\begin{pmatrix} \phi_{12}+\frac{2\pi R(1+\tau^{}_C)}{k}\\
e^{2\pi i/k}\,\phi_{34}\\
\end{pmatrix}.\label{eq:GactionZ3}
\end{equation}
Here $\phi_{12}$ and $\phi_{34}$ are the complex scalars which parametrize $C$ and $E$ respectively. Our focus here will be on $G=\mathbb{Z}_2$, where the action is the target space reflection on $E$ but translation on $C$. We first write down the partition function for $\mathrm{HS}_2$ and then explain the details:
\begin{equation}
\begin{split}
Z[\mathrm{HS}_2]&=\frac{1}{2}\Theta^{T^4}\left|\frac{\theta_1(z,\tau)}{\eta^3}\right|^4
+2\left|\frac{\theta_2(z,\tau)}{\theta_2(\tau)}\right|^2\cdot
\Theta^{T^2}_{\text{w/\ }G\text{-insertion}}\cdot
\left|\frac{\theta_1(z,\tau)}{\eta^3}\right|^2\\
&\ \ +2\left|\frac{\theta_4(z,\tau)}{\theta_4(\tau)}\right|^2\cdot
\Theta^{T^2}_{\text{w/\ }1/2-\text{shifted lattice}}\cdot
\left|\frac{\theta_1(z,\tau)}{\eta^3}\right|^2\\
&\ \ +2\left|\frac{\theta_3(z,\tau)}{\theta_3(\tau)}\right|^2\cdot
\Theta^{T^2}_{\text{w/\ }G\text{-insertion, }1/2-\text{shifted
lattice}}\cdot \left|\frac{\theta_1(z,\tau)}{\eta^3}\right|^2.\label{eq:Zhs22}
\end{split}
\end{equation}
The partition function is derived in Appendix B of \cite{ArabiArdehali:2023kar}. The Jacobi theta functions and Dedekind eta function in the above expression come from the oscillator modes of compact bosons and their superpartner fermions. The topological contributions, which are central to this paper, will be elaborated on below.\\

Let's start with the first line of \eqref{eq:Zhs22}. The first line is the contribution from the untwisted sector of the $\mathbb{Z}_2$ orbifold. The first topological term is a standard $T^4$ contribution\footnote{This convention is a little different from that of the Yellow Book \cite{DiFrancesco:1997nk}. One can go from the Yellow Book convention to ours by a simple rescaling $\sqrt{2} R_{yb}= R_i$.}
\begin{eqnarray}
    \Theta^{T^4}= \sum_{m_i,w_i} q^{\frac{\sum_i(\frac{m_i}{R_i}+w_i R_i)^2}{4}} \bar {q}^{  \frac{\sum_i(\frac{m_i}{R_i}-w_i R_i)^2}{4}}.
\end{eqnarray} 
The contribution is written as the sum over $m_i,w_i$, where $i=1,2,3,4$ labels the four cycles of the $T^4$. Here $m_i,w_i$ are the vertex and vortex quantum numbers (or momentum and winding numbers) for each $S^1$ inside $T^4$.\\
  
Let us now discuss the second topological piece $\Theta^{T^2}_{\text{w/\ }G\text{-insertion}}$. This comes from the untwisted sector but with the non-trivial element $G$ of $\mathbb{Z}_2$ inserted in the trace, as is standard in orbifold calculations, where the orbifold partition function is evaluated as
\begin{equation}
    \mathrm{Tr}_{\mathrm{orb}}=\frac{1}{2}\mathrm{Tr}_+(1+G)+\frac{1}{2}\mathrm{Tr}_-(1+G).\label{eq:orbTrZ2}
\end{equation}
Here $+$/$-$ indicates the sectors with untwisted/twisted boundary conditions along the spatial circle. The first factor $(1+G)/2$ projects onto the $G$-singlets. The second factor $(1+G)/2$ also projects to the singlets, but in the twisted Hilbert space.\\

We can understand the topological $\Theta$ contributions from the orbifold picture. The compact scalar which is $2 \pi R$ periodic, upon the action of $G$ becomes $\pi R$ periodic. The group acts as  $G:\phi\mapsto\phi+\pi R$. This action simply changes the radius from $R$ to $R/2$. Therefore we simply find the partition function of the orbifolded bosons by replacing $R$ with $R/2$ in the standard result known in CFT literature (for instance Eq.~(10.61) in \cite{DiFrancesco:1997nk}). But we will decompose this orbifold partition function as a trace as emphasized in (\ref{eq:orbTrZ2}), having untwisted and twisted sectors. Consider the partition of the compact boson with radius $R/2$:
\begin{equation}
\begin{split}
    Z(R/2)&=\frac{1}{2}\left(R  Z_{\mathrm{bos}}(\tau)\sum_{m,m'\in\mathbb{Z}}\mathrm{exp}-\frac{2 \pi R^2|(m/2)\tau-m'/2|^2}{2\mathrm{Im}\tau}\right)\\
    &=\frac{1}{2}\left((m\mathrm{\ even},m'\mathrm{\ even})+(m\mathrm{\ even},m'\mathrm{\ odd})+(m\mathrm{\ odd},m'\mathrm{\ even})+(m\mathrm{\ odd},m'\mathrm{\ odd})\right).\label{eq:ZR2}
\end{split}
\end{equation}
In the second line, we have split the sum into four sectors. The interpretation of the four terms aligns exactly with the four terms written in \eqref{eq:orbTrZ2}. With the expression $Z_{\mathrm{bos}}(\tau)=\frac{1}{\sqrt{\mathrm{Im}\tau}|\eta(\tau)|^2}$, the topological contributions $\Theta^{S^1}$ are defined as the factor multiplying $\frac{1}{|\eta(\tau)|^2}$ in first term $(m\mathrm{\ even},m'\mathrm{\ even})$ in the second line of \eqref{eq:ZR2}. We have
\begin{equation}
    \Theta^{S^1}=\frac{R}{\sqrt{\mathrm{Im}\tau}}\sum_{\substack{m\in\mathbb{Z}\\ m' \in\mathbb{Z}}}\mathrm{exp}-\frac{2 \pi R^2|m\tau-m'|^2}{2\mathrm{Im}\tau}.
\end{equation}
Here we have summed over $m,m'\in\mathbb{Z}$ by replacing $m/2\to m,$ and $m'/2\to m'$.  Then using the familiar Poisson resummation over $m'$ in the sum, we get
\begin{equation}
    \Theta^{S^1}=\sum_{\substack{m\in\mathbb{Z}\\ e \in\mathbb{Z}}}q^{(e/R+mR)^2/4}\bar{q}^{(e/R-mR)^2/4}.\label{eq:ThetaS1}
\end{equation}
Indeed we have gotten the expected 1st term on the RHS of \eqref{eq:orbTrZ2} from the 1st term on the second line of \eqref{eq:ZR2}.

Now, the second term with $(m\mathrm{\ even},m'\mathrm{\ odd})$ in the sum on the second line in (\ref{eq:ZR2}) gives
\begin{equation}
    \Theta^{S^1}_{\text{w/\ }G\text{-insertion}}=\frac{R}{\sqrt{\mathrm{Im}\tau}}\sum_{\substack{m\in\mathbb{Z}\\ m' \text{\ odd}}}\mathrm{exp}-\frac{2\pi R^2|m\tau-m'/2|^2}{2\mathrm{Im}\tau}.\label{eq:ThetaS1wG}
\end{equation}
Here we have used the $m\mathrm{\ even}$ constraint to redefine $m/2\rightarrow m$ and sum over all integers. For  $m'\text{\ odd}$ we write $\sum_{m'\text{\ odd}}=\sum_{m'}-\sum_{m'\text{\ even}}$. Then we use Poisson resummation for the $m'$ sum in (\ref{eq:ThetaS1wG}), which gives 
\begin{equation}
    \Theta^{S^1}_{\text{w/\ }G\text{-insertion}}=\sum_{m\in\mathbb{Z}}e^{-4\pi R^2 m^2\tau_2/4}\left(\sum_{e\in\mathbb{Z}}e^{-\frac{4\pi}{a}(e^2+\frac{2be}{4\pi i})}-\sum_{\tilde{e}\text{\ odd}}e^{-\frac{\pi}{a}(\tilde{e}^2+\frac{2b\tilde{e}}{2\pi i})}\right).\label{eq:ThetaS1wG2}
\end{equation}
Here $\tau=\tau_1+i\tau_2$, $a=2R^2/2\tau_2$, and $b=2\pi m R^2\tau_1/\tau_2$. Upon further simplification we get
\begin{equation}
    \Theta^{S^1}_{\text{w/\ }G\text{-insertion}}=\sum_{m\in\mathbb{Z}}\big(\sum_{e\ \mathrm{even}}-\sum_{e\ \mathrm{odd}}\big)\ q^{(e/R+mR)^2/4}\bar{q}^{(e/R-mR)^2/4}.\label{eq:ThetaS1wGf}
\end{equation}
This formula is also consistent with our expectation that  when $G$ acts on states of $e$ even gives $+1$, and when acting on the states having odd $e$ it gives $-1$.\\

The third and fourth terms in the sum of (\ref{eq:ZR2}) can be analyzed with similar manipulations. For the third term we have $m \in \mathrm{odd},m' \in \mathrm{even}$ in the sum. For the even $m'$ we will replace $m'/2\rightarrow m'$ and do the sum over all the integers. For the odd $m$ we write
\begin{equation}
\begin{split}
   \Theta^{S^1}_{\text{w/ $\frac{1}{2}$-shifted lattice}}=\frac{R}{\sqrt{\mathrm{Im}\tau}}\sum_{m \in \mathrm{odd},m'\in\mathbb{Z}}\left(\mathrm{exp}-\frac{2\pi R^2|(m/2)\tau-m'|^2}{2\mathrm{Im}\tau}\right).\\
\end{split}
\end{equation}
Then we can do the sum over $m' \in \mathbb{Z}$ using the Poisson resummation formula. We get
\begin{equation}
    \Theta^{S^1}_{\text{w/ $\frac{1}{2}$-shifted lattice}}=\sum_{\substack{m\in\text{odd}\\ e \in\mathbb{Z}}}q^{(e/R+mR/2)^2/4}\bar{q}^{(e/R-mR/2)^2/4}.\label{eq:ThetaS1shifted}
\end{equation}
Here $m \in \mathrm{odd}$ implies $m/2\in\mathbb{Z}+1/2$, justifying the name `$\frac{1}{2}$-shifted lattice'.\\

Finally, we analyze the fourth term in  (\ref{eq:ZR2}) having both $m$ and $m'$ odd. By following a similar reasoning as above we get
\begin{equation}
    \Theta^{S^1}_{\text{w/\ }G\text{-insertion, }1/2\text{-shifted
lattice}}=\sum_{m\in\mathrm{odd}}\big(\sum_{e\ \mathrm{even}}-\sum_{e\ \mathrm{odd}}\big)\ q^{(e/R+mR/2)^2/4}\bar{q}^{(e/R-mR/2)^2/4}.
\end{equation}
This term gives a contribution in the orbifold language as a twisted sector with $G$ inserted. This justifies the labeling of the lattice in the above equation. \\

This concludes our analysis for the $\mathbb{Z}_2$ shift orbifold of $S^1$. We need the orbifold of $\phi_{12}= T^2$ as written in \eqref{eq:GactionZ3}. The orbifold action suggests that the two $S^1$ of $T^2$ (labeled there as $\phi_{12}$) are independent and $g$ acts independently on both $S^1$. Hence, the orbifold of the $T^2$ or more precisely of $\phi_{12}$ are just the product of the results for the orbifolds of the two $S^1$. In summary, the topological contributions can be summarized as
 \begin{eqnarray}
&&\Theta^{T^2}=\sum_{\substack{m_i\in\mathbb{Z}\\ e_i \in\mathbb{Z}}}\prod_{i=1}^2 q^{(e_i/R_i+m_iR_i)^2/4}\bar{q}^{(e_i/R_i-m_iR_i)^2/4}, \\
 &&\Theta^{T^2}_{\text{w/\ }G\text{-insertion}}=\sum_{m_i\in\mathbb{Z}}\big(\sum_{e_i\ \mathrm{even}}-\sum_{e_i\ \mathrm{odd}}\big)\ \prod_{i=1}^2 q^{(e_i/R_i+m_iR_i)^2/4}\bar{q}^{(e_i/R_i-m_iR_i)^2/4},\\
   &&\Theta^{T^2}_{\text{w/ $\frac{1}{2}$-shifted lattice}}=\sum_{\substack{m_i\in\text{odd}\\ e_i \in\mathbb{Z}}} \prod_{i=1}^2 q^{(e_i/R_i+m_iR_i/2)^2/4}\bar{q}^{(e_i/R_i-m_iR_i/2)^2/4},\\
     &&\Theta^{T^2}_{\text{w/\ }G\text{-insertion, }\text{$\frac{1}{2}$-shifted
lattice}}=\!\!\sum_{m_i\in\mathrm{odd}}\big(\!\!\sum_{e_i\, \mathrm{even}}-\!\!\sum_{e_i\, \mathrm{odd}}\!\big)\ \prod_{i=1}^2 q^{(e_i/R_i+m_iR_i/2)^2/4}\bar{q}^{(e_i/R_i-m_iR_i/2)^2/4}.\nonumber\\
 \end{eqnarray}
The above sums are over $e_i,m_i, \text{with}\, i=1,2$. Here $i$ represents the two circles in $C=T^2$.

In \eqref{eq:Zhs22}, each of the four terms in \eqref{eq:orbTrZ2} are included as a product of a contribution from $C$ containing topological sectors as explained above, and a contribution from $E$ which contains topological contributions only through the first term in \eqref{eq:orbTrZ2} (see \emph{e.g.}~\cite{DiFrancesco:1997nk}).
This concludes our discussion of the $\mathrm{HS}_2$ partition function with the inclusion of topological sectors.

\bibliographystyle{JHEP}

\bibliography{refs}

\end{document}